\documentclass[aps,pre,twocolumn,groupedaddress,showkeys]{revtex4-1}
\usepackage{graphicx}
\usepackage{epstopdf}
\usepackage{amsmath}
\usepackage{multirow}
\usepackage{caption}
\usepackage{subcaption}

\begin{document}
\title{Pricing Weakly Model Dependent Barrier Products}
\author{Jan Kuklinski}
\email{jan.kuklinski@unil.ch}
\author{Panagiotis Papaioannou}
\author{Kevin Tyloo}
\affiliation{Facult\'e des HEC, Universit\'e de Lausanne, CH-1015 Lausanne, Switzerland}
\date{\today}


\begin{abstract}
We discuss the pricing methodology for Bonus Certificates and Barrier Reverse-Convertible Structured Products. Pricing for a European barrier condition is straightforward for products of both types and depends on an efficient interpolation of observed market option pricing. Pricing products with an American barrier condition requires stochastic modelling. We show that for typical market parameters, this stochastic pricing problem can be systematically reduced to evaluating only one fairly simple stochastic parameter being the asymmetry of hitting the barrier. Eventually, pricing Bonus Certificates and Barrier Reverse Convertibles with an American barrier condition, shows to be dependent on stochastic modelling only within a range of $\pm\frac{2}{3}$ of accuracy - e.g. within this accuracy limitation we can price these products without stochastic modelling. We show that the remaining price component is weakly dependent on the stochastic models. Combining these together, we prove to have established an almost model independent pricing procedure for Bonus Certificates and Barrier Reverse-Convertible Structured Products with American barrier conditions.
\end{abstract}

\maketitle

\section{Introduction}

Structured Products constitute a large part of the modern financial world. The way that those products are structured gives the opportunity for investors to have access to complex derivatives on a large variety of underlying assets. Due to the fact that they are pre-packaged investments they can be created in such a way that they satisfy varying risk/reward preferences. 

In their effort to make the product more attractive for the investors, the issuers often introduce some additional features the existence of which transforms the payoff of the instruments. One very popular feature is the introduction of a Barrier level. Depending on the structure of each product the role of the Barrier can be different. In any case, the introduction of a Barrier makes the pricing exEBRCise of such products much more challenging.

In this study we focus on the pricing of the Bonus Certificate and on the Barrier Reverse Convertible Product. The Bonus Certificate is attracting investors as it offer an have unlimited participation on the upside of the underlying asset together with some kind of protection depending on the event of the underlying hitting the Barrier level. 

The Barrier Reverse Convertible is offering a yield-enhanced profile as long as the Barrier level is not breached. This yield-enhancing property is very interesting, especially in the current environment characterized by very low return rates on conventional investment instruments.

For both instruments, breaching the barrier level is drastically affecting the performance. For European instruments, the breaching event occurs at the end of the life-cycle and calculating the probability of such a breach can be done using a probability distribution derived from interpolated option prices. Moreover, the valuation of a European product is functional on such a probability distribution and no path-dependent stochastic calculations are needed to perform the valuation.

For American instruments, the breaching event occurs whenever the price crosses the barrier level during the life-cycle of the product. Calculating the probability of such a breach requires path dependent stochastic calculus. At the first look, Bonus Certificates and Barrier Reverse Convertibles are strongly path dependent which means that a profound stochastic calculation is required to properly value these products.
Such a strong path-dependent calculation leads to two potentially difficult issues:
   
\begin{itemize}

\item [(a)] The choice of a stochastic model is not free of ambiguity and the final choice may strongly affect the result which is a non-desired feature.

\item [(b)] Assuming that a stochastic model is selected, it needs to be calibrated to option market data and resolved via Monte-Carlo calculation where both processes potentially complex and error-prone.

\end{itemize}
    
The central result of this paper is that we show that for typical market configurations relating to the market volatility levels and the barrier levels, stochastic modelization can be drastically simplified and the valuation of both the American Bonus Certificate (\textit{ABC}) and the American Barrier Reverse Convertible (\textit{ABRC}) depends only on one path-dependent parameter - the ratio of the amount of trajectories that have hit the barrier and ended above the barrier relative to the number of trajectories that have hit the barrier and ended below the barrier. 

As shown in this paper, the quite typical condition of separation between the barrier and the strike described as $B \ll S_{0}<K$ allows, with good approximation, to price the  \textit{ABC} contract

\begin{equation}
\label{first_equ}
\Pi_{ABC}^{model} \cong K - (K-B)(p_{H+}^{model} + p_{H-}) + Call(K, S_{0})
\end{equation}

In the above equation (\ref{first_equ}), only $p_{H+}^{model}$ (the probability of hitting the barrier and ending above and $p_{H-}$ is the same but ending below) is depending on a stochastic model e.g. is path dependent. The remaining parameters $(p_{H-}, Call(K, S_{0})$ are to be calculated from an interpolation of option price data. We rewrite this result as 

\begin{equation}
\Pi_{ABC}^{model} \cong K - (K-B)(2 + \delta^{model})p_{H-} + Call(K, S_{0})
\end{equation}
with $\delta^{model}=\frac{p_{H+}^{model}-p_{H-}}{p_{H-}}$. Taking $\delta^{model} = 0$ the basic ``model-free" approximation. Practical examples show that this gives the correct market price with an accuracy of $\pm\frac{2}{3}\%$. 

Accounting for the ``correct" value of $\delta^{model}$ requires using a specific stochastic model calibrated to the observed Volatility surface. As discussed in this paper, using two models with a quite different logic, gives a difference of no more that $\pm0.5\%$. This leads us to the conclusion that pricing the American Bonus Certificate goes via a very efficient and weakly model dependent modelization. As shown separately, the same logic applies to the American Barrier Reverse Convertible Product.

\section{The Bonus Certificate product with a single underlying}

Bonus Certificate contracts (\textit{BC} contracts) are based on the barrier payment rule described on the picture below:
 
\begin{figure}[h!]
\centering
\includegraphics[scale=1]{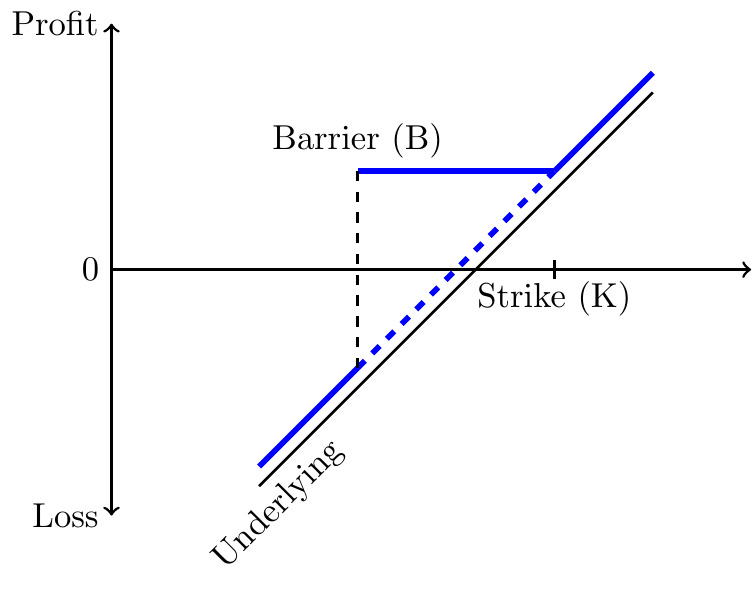}
\caption{\label{BC} Bonus Certificate payoff.}
\end{figure}

We shall focus on \textit{C} contracts that are based on a single underlying. In such a case the payment mechanism is typically described by two parameters; the Barrier \textit{B} and the Strike \textit{K}. 

For the  contract with European Barrier condition (\textit{EBC}), depending on the value of the underlying at expiry, the payoff of the contract is as follows:

\begin{itemize}

\item [(a)] if $S_{T}\geq B$  then the contract pays at the end $C_{0}=\max(S_{T}, K)$.
\item [(b)] if $S_{T}<B$ then the contract pays at the end $C_{1}=S_{T}$.

\end{itemize}

As seen from the above, as long the barrier is not breached at the end, the contract pays higher than its initial value.

For the American Bonus Certificate product (\textit{ABC}), the mechanism is different. If during the contract duration the barrier is not breached e.g. $S_{0<t<T}>B$, the contract pays at the end the following amount:

\begin{eqnarray}
C_{0} = \max(S_{T}, K)
\end{eqnarray}

If the barrier has been breached e.g. for some $t'$ we had $S_{t'}\leq B$, the payoff is proportional and no longer protected by the barrier:

\begin{eqnarray}
C_{1} = S_{T}
\end{eqnarray}

Such contracts are available on the market for a variety of underlyings. The typical parameters are such that the barrier and strike level are about 70\% and 105\% receptively of the initial spot price.

\subsection{Pricing the European Bonus Certificate}

We shall be pricing the \textit{EBC} contracts using interpolated probabilities described in Appendix \ref{appendixA}. 

For the \textit{EBC} the probability distribution of the underlying at expiry is determining the value of the product. Eventually the expected payoff of the \textit{EBC} contract is:

\begin{eqnarray}
\Pi_{EBC}^{interpolated} &=& \Pi_{BN}^{a} + \Pi_{BN}^{b} + \Pi_{BH} \\
\Pi_{BN}^{a} &=& K \int_{B}^{K}d\phi\, \rho_{T}^{interpolated}(\phi) \\
\Pi_{BN}^{b} &=& K + \int_{K}^{\infty} d\phi\, (\phi-K) \rho_{T}^{interpolated}(\phi) \\ \Pi_{BH} &=& \int_{-\infty}^{B} d\phi\,\phi \rho_{T}^{interpolated}(\phi)
\end{eqnarray}

As we see, the price constitute of three terms:

\begin{itemize}

\item [(a)] for $B\leq S_{T} \leq K$  the the payoff is $C_{0}=K$
\item [(b)] for $S_{T}>K$ we get $\Pi_{BN}^{b}$
\item [(c)] for $S_{T}<B$ we get $\Pi_{BH}$.

\end{itemize}

We can rewrite the payoff of the \textit{EBC} contract as 

\begin{eqnarray}
\Pi_{EBC}^{interpolated} &=& \Pi_{BN}^{\alpha} + \Pi_{BN}^{\beta} + \Pi_{BH}^{\alpha} - \Pi_{BH}^{\beta} \\
\Pi_{BN}^{\alpha} &=& K \int_{B}^{\infty}d\phi\, \rho_{T}^{interpolated}(\phi) \\
\Pi_{BN}^{\beta} &=& \int_{K}^{\infty} d\phi\, (\phi-K) \rho_{T}^{interpolated}(\phi) \\ \Pi_{BH}^{\alpha} &=& B\int_{-\infty}^{B} d\phi\, \rho_{T}^{interpolated}(\phi) \\
\Pi_{BH}^{\beta} &=& \int_{-\infty}^{B} d\phi\,(B-\phi) \rho_{T}^{interpolated}(\phi) 
\end{eqnarray}

The above formulae uses: 

\begin{eqnarray}
Call(K, S_{0}) &=& \int_{K}^{\infty} d\phi\, (\phi-K) \rho_{T}^{interpolated}(\phi) \\
Put(B, S_{0}) &=& \int_{-\infty}^{B}d\phi\,(B-\phi) \rho_{T}^{interpolated}(\phi) \\
p_{H-} &=& \int_{-\infty}^{B}d\phi\,\phi \rho_{T}^{interpolated}(\phi) 
\end{eqnarray}

Eventually the payoff of the \textit{EBC} contract is:

\begin{equation}
\label{EBC_int}
\Pi_{EBC}^{interpolated} = K - (K-B) p_{H-} + Call(K, S_{0}) - Put(B, S_{0})
\end{equation}

All of the quantities involved above require calculating the probability density $\rho_{T}^{interpolated}(\phi)$. The latter is done using the formulas:

\begin{eqnarray}
\rho_{T}^{interpolated}(\phi\leq S_{0}) &=& \frac{\partial^{2}}{\partial\phi^{2}}Put_{T}^{interpolated}(\phi\leq S_{0})\nonumber\\
\label{rho_interp_call}
\\
\rho_{T}^{interpolated}(\phi> S_{0}) &=& \frac{\partial^{2}}{\partial\phi^{2}}Call_{T}^{interpolated}(\phi> S_{0})\nonumber\\
\label{rho_interp_put}
\end{eqnarray}

The expression for the interpolated option prices adjusted to the observed market data are described in Appendix \ref{appendixA} together with the fitting procedure. The interpolation is based on a two-step adjustment:

\begin{itemize}
\item [(a)] The market prices are fitted to a formula extending on the option prices following the SABR stochastic model
\item [(b)] An adjustment is made for the tail prices
\end{itemize}

The analysis of the correspondence between market data and fitted prices is discussed in Chapter 5.

\subsection{Pricing the American Bonus Certificate product}

Following on the conditional probabilities defined in Appendix \ref{appendixB}, the expected payoff of the \textit{ABC} contract is:

\begin{eqnarray}
\Pi_{ABC}^{model} &=& \Pi_{BN}^{a} + \Pi_{BN}^{b} + \Pi_{BH} \\
\Pi_{BN}^{aa} &=& K \int_{B}^{\infty}d\phi\, \rho_{BN}^{model}(\phi) \\
\Pi_{BN}^{bb} &=& \int_{K}^{\infty} d\phi\, (\phi-K) \rho_{BN}^{model}(\phi) \label{ABC_b}\\ 
\Pi_{BH} &=& \int_{-\infty}^{\infty} d\phi\, \phi\rho_{BH}^{model}(\phi)
\end{eqnarray}

In the above we use the probability density conditional on the barrier not being hit ($\rho_{BN}^{model}(\phi)$) and conditional on the barrier being hit ($\rho_{BH}^{model}(\phi)$).  We can resolve immediately:

\begin{equation}
\rho_{BH}^{interpolated}(\phi < B) = \frac{\partial^{2}}{\partial\phi^{2}}Put_{T}^{interpolated}(\phi\leq B)
\end{equation}

The above follows from the fact that if $S_{T}<B$ the barrier must have been hit so $\rho_{BH}(\phi<B)=\rho_{T}^{interpolated}(\phi\leq B)$. Separately if the barrier has been hit before maturity, the martingale property underlying the stochastic models leads to:

\begin{eqnarray}
\int_{-\infty}^{\infty}d\phi\, \phi \rho_{BH}^{model}(\phi) = B
\end{eqnarray}

We further use an approximation for the part given in (\ref{ABC_b}) (the approximation consist of neglecting the contribution from trajectories that hit the barrier and eventually ended above the strike -- see Appendix \ref{appendixB}:

\begin{eqnarray}
\Pi_{BN}^{bb} & = &\int_{K}^{\infty} d\phi\, (\phi-K) \rho_{BN}^{model}(\phi) \\
& \cong & \int_{K}^{\infty} d\phi\, (\phi-K) \Big(\rho_{BN}^{model}(\phi) + \rho_{BH}^{model}(\phi) \Big)\\
& = & \int_{K}^{\infty} d\phi\, (\phi-K) \rho_{T}^{model}(\phi)  = Call(K, S_{0})
\end{eqnarray}

We define the model independent $p_{H-}$ (breach-and-down) together with the model dependent $p_{H+}^{model}$ (breach-and-up) probabilities:

\begin{eqnarray}
p_{H-} &=& \int_{-\infty}^{B}d\phi\,\rho_{T}^{interpolated}(\phi) \\
p_{H+}^{interpolated} &=& \int_{B}^{\infty} d\phi\,\rho_{BH}^{model}(\phi)
\end{eqnarray}

The probability $p_{H-}$ corresponding to breaching the barrier $S_{t}<B$ and ending below the barrier at termination ($S_{T}<B$) is path-independent and is to be calculated from the probability distribution of the underlying at expiration $\rho_{T}^{interpolated}(\phi)$.

The probability $p_{H+}^{model}$ corresponding to breaching the barrier $S_{t}<B$ and ending above the barrier at termination ($S_{T}>B$) is path-dependent and cannot be deduced solely from the distribution $\rho_{T}^{interpolated}(\phi)$. 

Eventually we combine the two and obtain the expression for the price of the American Bonus Certificate:

\begin{equation}
\Pi_{ABC}^{model} \cong K - (K-B)(2+\delta^{model})p_{H-} + Call(K, S_{0})
\end{equation}

With the $\delta^{model}$ parameter:

\begin{equation}
\delta^{model} = \frac{p_{H+}^{model}-p_{H-}}{p_{H-}}
\end{equation}

The zero-level result is obtained for $\delta^{model}=0$ which corresponds to using a Volatility Surface without skew. The parameters ($p_{H-}$, $Call(K, S_{0})$) are to be calculated form interpolated option data which is a model independent procedure. As shown in Chapter 5, analyzing market data shows that this crude approximation gives an accuracy of $\pm\frac{2}{3}\%$. This result is contrary to the common belief, that a detailed stochastic calculus, typically basing on a Monte-Carlo calculation, is needed to calculate the valuation of the American Bonus Certificate (the same result applies to the American Barrier Reverse Convertible).

To get a further insight, the parameter $\delta^{model}$  needs to be calculated using a stochastic model. It is functional on the amount of trajectories that have hit the barrier before maturity and have ended above the barrier. In Appendix \ref{appendixB} we have calculated $\delta^{model}$ calibrated to two market data sets using two different stochastic models. The conclusion is that the variation of $\delta^{model}$ is affecting the valuation of the \textit{ABC} product by less than $\pm1\%$. Such conclusions are not general but they reflect the typical relation between market volatility levels and the barriers levels.

\section{Pricing the mono Barrier Reverse Convertible product}

The Barrier Reverse Convertible contracts (\textit{RC} contracts) are based on the barrier payment rule described on the picture below (we consider again only single/mono-underlying products:

\begin{figure}[h!]
\centering
\includegraphics[scale=1]{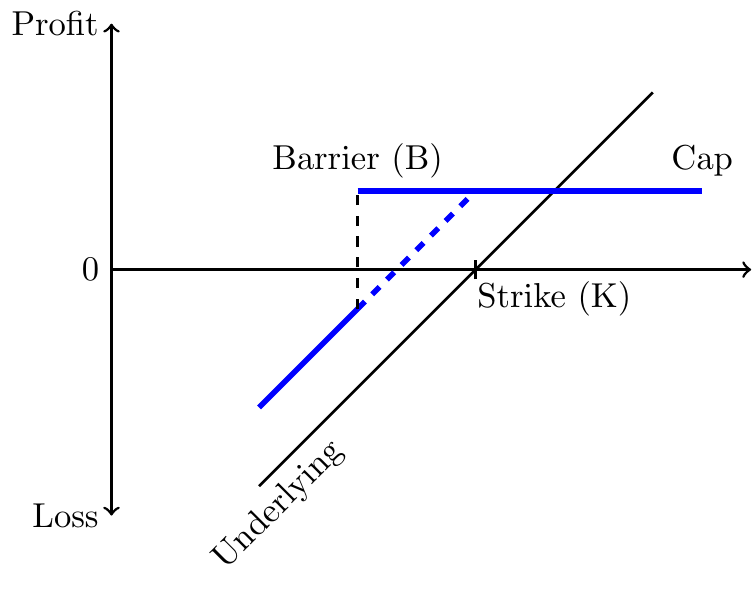}
\caption{\label{fig_BRC} Barrier Reverse Convertible payoff.}
\end{figure}

We shall focus on \textit{BRC} contracts based on a single underlying with the following relation between the coupon ($C_{0}=S_{0}+R$) and the barrier level (\textit{B}): $B\leq \frac{3}{4}C_{0}$.

\subsection{Pricing the European mono Barrier Reverse Convertible}

A European Barrier Reverse Convertible contracts (\textit{EBRC}) pays a yield-enhanced payoff $C_{0}=S_{0}+R$ if at contract maturity the underlying is above the barrier level ($S_{T}\geq B$). In the opposite case ($S_{T}<B$), the contract pays at maturity $C_{1}=S_{T}$, which means that the loss is potentially unlimited.

Pricing the European Barrier Reverse Convertible contract goes along the same lines as for the  \textit{EBC} contract - we integrate the probability distribution obtained from interpolating option prices the contract payoff:

\begin{eqnarray}
\Pi_{EBRC}^{interpolated} &=& \Pi_{BN} + \Pi_{BH}^{a} - \Pi_{BH}^{b} \\
\Pi_{BN} &=& (S_{0} + R) \int_{B}^{\infty}d\phi\,\rho_{T}^{interpolated}(\phi) \\
\Pi_{BH}^{a} &=& B \int_{-\infty}^{B}d\phi\, \rho_{T}^{interpolated}(\phi) \\
\Pi_{BH}^{b} &=& \int_{-\infty}^{B}d\phi\, (B-\phi) \rho_{T}^{interpolated}(\phi)
\end{eqnarray}

Eventually the \textit{EBRC} price has a simple form:

\begin{equation}
\Pi_{EBRC}^{interpolated} = S_{0} + R - (S_{0} + R - B)p_{H-} - Put(B, S_{0})
\end{equation}

\subsection{Pricing the American mono Barrier Reverse Convertible}

An American Barrier Reverse Convertible contracts (\textit{ABRC}) pays a yield-enhanced payoff $C_{0} = S_{0} + R$  if at any time  $t$  during the contract duration the barrier level was not breached e.g.  $S(0<t\leq T)\geq B$ . In the opposite case $S_{T}<B$ , the contract pays at maturity $C_{1} = S_{T}$, which again means that the loss is potentially unlimited. 

Pricing the American Barrier Reverse Convertible requires using the conditional probabilities

\begin{eqnarray}
\Pi_{ABRC}^{model} &=& \Pi_{BN} + \Pi_{BH}^{a} + \Pi_{BH}^{a} \\
\Pi_{BN} &=& (S_{0} + R) \int_{B}^{\infty}d\phi\,\rho_{BN}^{model}(\phi) \\
\Pi_{BH}^{a} &=& \int_{-\infty}^{S_{0}+R}d\phi\, \phi \rho_{BH}^{model}(\phi) \\
\Pi_{BH}^{b} &=& (S_{0} + R)\int_{S_{0}+R}^{\infty}d\phi\,\rho_{BH}^{model}(\phi)
\end{eqnarray}

Again we calculate the breach-and-down probability $p_{H-}$  and the breach-and-up probability $p_{H+}^{model}$. We also use an approximation following the assumptions involved in deriving Eq.(\ref{ABC_b}) (see Appendix \ref{appendixB}):

\begin{eqnarray}
\Pi_{BH}^{a} &=& \int_{-\infty}^{S_{0}+R}d\phi\, \phi \rho_{BH}^{model}(\phi)\\
& \cong &\int_{-\infty}^{\infty}d\phi\, \phi \rho_{BH}^{model}(\phi) = B \\
\int_{S_{0}+R}^{\infty}d\phi\,\rho_{BH}^{model}(\phi) &\cong &0 \longrightarrow \Pi_{BH}	^{b} \cong 0
\end{eqnarray}

Eventually the \textit{ABRC} price has the form:

\begin{equation}
\Pi_{ABRC}^{model} \cong S_{0} + R - (S_{0} + R - B) (2 + \delta^{model})p_{H-}
\end{equation}

As expected, the form of the \textit{ABRC} price is very similar to the \textit{ABC} price. Again for typical values of the market volatility levels and for barriers corresponding to 60\%-70\% of the underlying level, we can calculate the value of $\Pi_{ABRC}^{model}$ using $\delta^{model}=0$ and this will lead to an accuracy not worse than $\pm3\%$ . Resolving for a better pricing accuracy requires using a stochastic calculation for $\delta^{model} \neq 0$. While the calculation of $\delta^{model}$ requires calibration to the Volatility Surface and is model sensitive, the sets of market parameters that we have discussed in Appendix \ref{appendixC} leads to a conclusion that a price accuracy of $\pm1\%$ can be obtained and this result includes discrepancies related to choosing alternative stochastic models.

\section{Analysing market data}

\subsection{Product Selection}

For the purpose of this study we have chosen to price two Bonus Certificates which are listed on the Swiss market (SIX Swiss Exchange AG). Both of them are single underlying products. The underlying of the first product (with ISIN CH0245337099) is the S\&P 500 Index while the second one (with ISIN CH0245336034) is written on the EURO STOXX50 Price Index. The first instrument has the barrier observation only at maturity (European Bonus Certificate) whereas the barrier observation for the second one is continuous (American Bonus Certificate). The reasoning behind the selection of those specific products has to do with the fact that in this study we are interested in examining the behaviour of structured products written on indices instead of individual stocks. That is because options on the main indices exhibit some standardized patterns. Moreover, due to the large demand and liquidity of such options the task of finding and extracting data of both good quality and sufficient quantity becomes much more feasible. Finally, we choose single underlying products and not Basket Bonus Certificates due to the additional difficulty that such products exhibit that is mainly created by the necessity of modeling the correlation between the underlyings included in the basket, a topic which is beyond the scope of this study.

\subsection{European Bonus Certificate}
\subsubsection{Data Selection}

We have chosen to price the European Bonus Certificate written on the S\&P 500 for seven different pricing dates starting on 02/02/2015 and ending on 30/11/2015 (See Appendix for detailed representation of datasets). The reason why we choose these dates is because the maturity of this product is up to three years which in turn means that for pricing purposes we need call and put prices on plain vanilla options with underlying asset the S\&P 500 Index. However, since we can not find quotes for options on S\&P 500 with maturity longer than 2.5 years we are obliged to set our first pricing date on 02/02/2015. Another reason why we have selected those specific days has to do with the variety of levels at which the product is priced on the selected dates. As it will be shown and discussed in further detail later on, the market prices for the chosen dates are quite different spanning from the level of \$189.7 to \$207.

For each of the above selected dates, we have decided to download prices of both call and put options for all strikes and all available maturities (longer than two months) on the S\&P 500. These prices are calibrated using the three different models described in the Appendix \ref{appendixA}. The calibration results are about to determine the model which we are going to use in order to price our product. Moreover, the calibrated parameters are essential to the pricing of the European Bonus Certificate since they enter (\ref{EBC_int}) through $p_{H-}$ which as mentioned in Chapter 2 is calculated using the cumulative density function obtained by the model at hand.

\subsubsection{Results}

We have chosen to price the European Bonus Certificate (\textit{EBC}) using the Static Shifted Log-Lognormal Model (SSLN) as well as the SABR ($\beta = 1$). As one can see from our calibration results, the $\rho$ parameter for the SABR ($\beta = 1$) model is very often equal to 0.99. In that case the SABR ($\beta = 1$) model becomes the so-called Exponential SLN. The method that we use consists of the following steps. First, we calibrate the observed market prices of the options for all the available dates. For the two dates, the one before and the one right after the product maturity date we interpolate the model generated prices after using the optimal calibrated parameters for each date. Second, we obtain our parameters corresponding to the desired date after fitting higher order polynomials for better accuracy. For more details regarding the polynomial fits and the interpolation results see Appendix \ref{appendixA}. Finally, after having a set of optimal and interpolated parameters for each pricing date we plug them into equation (\ref{EBC_int}) to get the model price.

As it can be seen form Table \ref{EBC_err} the prices obtained by both models are close to the market price. The SSLN seems to give an overpriced value of 2.07\% on average while the SABR ($\beta = 1$) is 1.87\% off on average (in absolute terms).This results seems to be consistent with our assumption that the prices obtained using either SSLN or SABR should not exhibit a significant difference since the fitting of the market prices around the barrier level of 79.5\% seems negligible (see figure with barrier).
\\

\begin{table}[h!]
\centering
\begin{tabular}{ l c c c c} 
Price Date & Market Price & SSLN Price & SABR$_{\beta = 1}$Price \\ 
\hline \\
02/02/2015 & 194.1 & 199.87 (2.97\%) 	&	195.99 (0.98\%) \\
24/03/2015 & 205.2 & 208.56 (1.64\%) 	&	204.11 (-0.53\%) \\
26/05/2015 & 206.1 & 209.28 (1.54\%)	&	202.51 (-1.75\%) \\
24/08/2015 & 189.7 & 189.77 (0.04\%) 	&	185.43 (-2.25\%) \\
22/09/2015 & 191.5 & 195.14 (1.90\%)	&	194.42 (1.52\%) \\
22/10/2015 & 202 & 205.03 (1.50\%) 	&	196.85 (-2.55\%) \\
30/11/2015 & 207 & 208.89 (0.92\%) 	&	199.73 (-3.51\%) \\
\end{tabular}
\caption{\label{EBC_err} European Bonus Certificate pricing results.The number in parenthesis is the error between the model and the market.}
\end{table}

\subsection{American Bonus Certificate}
\subsubsection{Data Selection}

For the case of the American Bonus Certificate we have chosen again seven different pricing dates starting on 08/07/2014 and ending on 01/09/2015. Since the maturity of this product is two years, it is possible to start pricing it at the initiation (first listing date) since we are able to obtain put and call prices on the EUROSTOXX 500 Price Index for maturities longer than two years. Again, as in the case of the European Bonus Certificate, the date selection is made in such a way that different levels in the lifetime of the product are reflected. The pricing range in that case spans from 90.2\% to 108.3\%. We follow exactly the same steps as in the case of the European product, obtaining call and put prices for all available strikes and maturities on the EUROSTOXX 500 and then calibrating them using
the models mentioned in the previous section.

\subsubsection{Results}

For the American Bonus Certificate we use the same procedure as in the European one, however we go one step further by making use of the Monte Carlo method which is necessary to compute the delta parameter that enters into equation (1.2). We perform Monte Carlo (MC) simulations using two different models, the Static $q$ SLN, the Dynamic $q$ SLN. For the dynamic $q$ case we choose to run our MC simulation after assuming that $q$ follows a power-law (see Appendix). However, as Table \ref{ABC_err}  shows, it is remarkable that the most accurate prices are obtained by the SLN after assuming that the $\delta$ parameter is equal to 0. More specifically, the average absolute difference between the model and the market prices are 1.08\%, 2.26\% and 2.37\% for SLN plus $\delta = 0$, SSLN MC, and Dynamic SLN MC respectively.

\begin{table*}[t]
\centering
\begin{tabular}{ l c c c c} 
Price Date & Market Price & SLN(plus $\delta=0$) & SLN Static MC &SLN Dynamic MC\\
\hline
\\
08/07/2014 & 97.6 & 94.82 (-2.85\%) 	&	94.37 (-3.30\%) 	& 94.12 (-3.56\%)	\\
08/08/2014 & 90.2 & 89.93 (-0.30\%) 	&	91.77 (1.74\%) 	& 91.66 (1.62\%)		\\
19/09/2014 & 96.8 & 97.20 (0.42\%)	&	96.16 (-0.66\%) 	& 96.01 (-0.81\%)	\\
28/11/2014 & 97 & 96.26 (-0.75\%) 	&	95.36 (-1.69\%) 	& 95.29 (-1.76\%)	\\
03/02/2015 & 102.1 & 99.81 (-2.24\%)	&	98.54 (-3.48\%) 	& 98.32 (-3.70\%)	\\
01/06/2015 & 108.3 & 108.77 (0.44\%)	&	106.57 (-1.59\%)	& 106.48 (-1.68\%)	\\
01/09/2015 & 96.65 & 97.20 (0.57\%)	&	99.90 (3.37\%) 	& 99.96 (-3.43\%)	\\
\end{tabular}
\caption{\label{ABC_err} American Bonus Certificate pricing results.The number in parenthesis is the error between the model and the market.}
\end{table*}

\section{Conclusion}

Bonus Certificates and Barrier Reverse Convertible contracts are popular Barrier Structured Products. Pricing these products for a European Barrier condition can be done providing option price data matching the expiry of the selected Structured Product are available. If so, a standard pricing methodology based on interpolating option prices can be applied. In this paper we have discussed an interpolating methodology based on extrapolating the SABR formulas. We eventually conclude, that for a wide range of market parameters, such an interpolation can be restrained to the most elementary level consisting of using Shifted-Log-Normal based pricing.

American Bonus Certificates (\textit{ABC}) and American Barrier Reverse Convertible (\textit{ABRC}) contracts are path sensitive products. Their pricing is typically regarded as requiring a full stochastic modelization requiring calibration to the entire Volatility surface and typically a Monte-Carlo calculation to find the correct pricing.

The data analysis presented herein is showing, that both the price of \textit{ABC} and  \textit{ABRC} products can be described as a sum on strongly decreasing terms. Surprisingly, for a quite wide range of market and product parameters, the leading terms which are carrying no less than 96\%-97\%  of the product value, are path-independent e.g. can be calculated without stochastic modelization. 

The expressions for the value of the  and of the  products have the following form:

\begin{eqnarray}
\Pi_{ABC}^{model} &=& K - (K-B)(2 + \delta^{model})p_{H-} + Call(K, S_{0}) \epsilon_{BC}\nonumber\\ \label{ABC_conc}\\
\Pi_{ABRC}^{model} &=& K - (K-B)(2 + \delta^{model})p_{H-} + \epsilon_{RC} \label{ABRC_conc}
\end{eqnarray}

In both prices the contribution form the correction terms ($\epsilon_{BC}$, $\epsilon_{RC}$) is affecting the price of less than $\pm0.5\%$. The quantities $p_{H-}$ and $Call(K, S-{0})$ are to be calculated from interpolated option at-maturity prices.

The linear terms proportional to $\delta^{model}$ are the only one which are reflecting the path-sensitive aspect of the prices (\ref{ABC_conc}) and (\ref{ABRC_conc}). The parameter  has a simple meaning being proportional probability of breaching the barrier and ending above it versus breaching the barrier and ending below it. This means that stochastic calculations can be reduced to calculating  only.

Stochastic calculation of requires selecting a stochastic model, calibrating it to the Volatility Surface and later on calculating the path dependent value. All these steps are prone to errors with an additional ambiguity of selecting the right stochastic model. Yet these potential errors and discrepancies are further ``compressed" as they contribute to a component that carries no more than $\frac{1}{30}$ of the total price. Eventually, for typical parameters of  and  products, we can determine its price with an accuracy of no less that $\pm1\%$ including discrepancies caused by using alternative stochastic models to calculate.

We conclude that in a typical scenario relating market volatilities and the parameters of a single-underlying product the following inequality applies:

\begin{eqnarray}
\label{condition_conc}
\frac{3}{2}\frac{K-B}{K} \geq \sigma_{1 \text{year}}^{ATM}\sqrt{\Delta T}
\end{eqnarray}

e.g. the barrier level follows:

\begin{eqnarray}
B \leq K \big( 1-\frac{2}{3}\sigma_{1 \text{year}}^{ATM}\sqrt{\Delta T} \big)
\end{eqnarray}

In the above equation (\ref{condition_conc}), $\sigma_{1 \, year}^{ATM}$ is the volatility of 1 year options on the underlying in question and $\Delta T$ is the duration of the product at initiation.

Assuming that (\ref{condition_conc}) is holding, we can make an approximation and disregard the density of trajectories that starts at, reaches the barrier at some stage and eventually ends above the strike. If this assumption is correct, pricing of the American Bonus Certificates and American Barrier Reverse Convertible contracts is greatly simplified and we end with the expressions (\ref{ABC_conc}) and (\ref{ABRC_conc}).

The result given by the $\delta$-expressions gives an insight into the influence of the skew and smile of the Volatility Surface on the product prices - namely all is reflected solely in the $\delta^{model}$ parameter and selecting $\delta^{model} = 0$  corresponds to using a symmetric diffusion model.

In summary we have shown that Bonus Certificates and Barrier Reverse Convertible contracts with both American and European barrier conditions can be priced with a very similar accuracy, while the pricing for American products has a relatively small path-dependent ingredient.

\nocite{SVSP, Gat, Dup, JRK_D, Kor, Lor1}
\bibliography{pricing_WMD_barrier_products_V3}{}
\bibliographystyle{plain}


\clearpage
\appendix 

\section{Pricing Vanilla products and interpolating market option prices}
\label{appendixA}

Vanilla Call and Put options are functional on the probability distribution (e.g. on the probability density $\rho(\phi)$) at expiration:

\begin{eqnarray}
Call(K, S_{0}, t, T) &=& V D_{t,T} \int_{K}^{\infty}d\phi\, (\phi-K) \rho({\phi}) \\
Put(K, S_{0}, t, T) &=& V D_{t,T} \int_{-\infty}^{K}d\phi\, (K-\phi) \rho({\phi})
\end{eqnarray}

Where $D_{t,T}$ is the discount factor and $V$ is the Volume factor. The above expressions can be extended to a Vanilla Structured Product (SP) with arbitrary payoff function $f(\phi)$: 

\begin{eqnarray}
\label{SP_app_A}
\Pi_{SP}^{Vanilla}(S_{0}, t, T) &=& V D_{t, T} \int_{-\infty}^{\infty} d\phi\, f(\phi) \rho(\phi)
\end{eqnarray}

The European Bonus Certificate and the European Barrier Reverse Convertible products fall in the category described by equ. (\ref{SP_app_A})

The distribution of the underlying has the general form $\rho(F_{t,T}, t, T; \phi)$ and its mean value is called the Forward $F_{t, T} = \int_{-\infty}^{\infty}d\phi\,\phi\rho(\phi)$.

The evolution of the probability distribution $\rho(F_{t,T}, t, T; \phi)$ is typically assumed to follows a martingale stochastic motion. 

The shape of the forward curve $F_{t, T}$ and the discounting function $D_{t,T}$ can be determined from market data and easily superimposed on the stochastic models that we are using. For the sake of simplicity, all formulas presented in this paper will use the approximation $F_{t, T}=S_{0}\equiv F$ and $D_{t,T} = 1$ together with $V=1$ the assumption, whereas in the calculations we use $V=1$ while the shape of the Forward and Discount curve are adjusted to market data.

We start with pricing Vanilla products expiring at $t = T$ under the assumption, that market prices of option expiring at the same date, or very close to it, are available. In such a case, basing on a selected stochastic or extrapolating model, we find the interpolated prices of Vanilla options. Using these prices, we calculate the cumulative (cdf) probability distribution:

\begin{eqnarray}
D_{T}^{interpolated}(\phi \leq S_{0}) &=& \frac{\partial}{\partial\phi} Put_{T}^{interpolated}(\phi \leq S_{0}) \\
D_{T}^{interpolated}(\phi > S_{0}) &=& 1 - \frac{\partial}{\partial\phi} Call_{T}^{interpolated}(\phi > S_{0})\nonumber\\
\end{eqnarray}

We shall use several models to interpolate observed option market prices. For all of them we adjust the model parameters using the least square principle and minimize the quantity $E^{2}$ following:

\begin{eqnarray}
E^{2} &=& \sum_{j=1}^{n} \Big (\frac{Put_{T}^{market}(\phi_{j} \leq S_{0}) - Put_{T}^{model}(\phi_{j})}{Put_{T}^{market}(\phi_{j} \leq S_{0}) + Put_{T}^{model}(\phi_{j})} \Big)^{2}\nonumber\\
& & +\sum_{i=1}^{n} \Big( \frac{Call_{T}^{market}(\phi_{i} > S_{0}) - Call_{T}^{model}(\phi_{i})}{Call_{T}^{market}(\phi_{i} > S_{0}) + Call_{T}^{model}(\phi_{i})} \Big)^{2}\nonumber\\
\end{eqnarray}

On that basis we calculate $\rho_{T}^{interpolated}(\phi)$ following the formulas (\ref{rho_interp_call}) and (\ref{rho_interp_put}).

For interpolating market data option prices, we use alternatively three models:

\begin{itemize}

\item [(A)] The Shifted Log-Normal Model

The SLN model follows:
\begin{equation}
D^{SLN}(K) = \mathrm{N}_{0}(-d_{2})
\end{equation}
\begin{eqnarray}
Call^{SLN} &=&DV\times\nonumber\\
& &\hspace{-0.8cm}  \Big( (F-K)\mathrm{N}_{0}(d_{2}) + \frac{F}{q} \big(\mathrm{N}_{0}(d_{1}) - \mathrm{N}_{0}(d_{2}) \big) \Big) \\
Put^{SLN} &=& DV \times\nonumber\\
& & \hspace{-1.6cm}\Big( (K-F)\mathrm{N}_{0}(-d_{2}) + \frac{F}{q} \big(\mathrm{N}_{0}(-d_{1}) - \mathrm{N}_{0}(-d_{2}) \big) \Big)
\end{eqnarray}

where $\bar{\sigma} = \sigma_{q}\sqrt{T-t}$, $d_{1,2} = -\frac{1}{q\bar{\sigma}}\ln\big(1 + q \frac{K-F}{F} \big) \pm \frac{1}{2}q\bar{\sigma}$.

The Shifted Log-Normal Model has two parameters to be adjusted to the market: $\bar{\sigma}$  and $s=q\bar{\sigma}$. The first parameter is almost proportional to the price of the at-the-money option:

\begin{eqnarray}
\Pi_{ATM}^{SLN} = \frac{FDV\bar{\sigma}}{s} \big( 2\mathrm{N}_{0}(\frac{s}{2})-1 \big) \cong \frac{FDV\bar{\sigma}}{\sqrt{2\pi}}
\end{eqnarray}

The second parameter $s$ is related to the near-at-the-money asymmetry of the call and put option prices. For $s=0$ we have ``no-skew", e.g. option prices obeys symmetric Bachelier pricing.

The SLN model is typically giving a good description of near-at-the-money options, while it often cannot reproduce prices for very distant strikes. On the findings of the analysis discussed in this paper, is that when analysing the features of barrier products on skewed markets, the market interpolation provided by the SLN model often proves satisfactory.

\item [(B)] The SABR pricing

The SABR stochastic \cite{HagLes, JanObl} model is very widely used in option pricing. The approximated yet typically accurate expressions for option prices are given through strike dependent Implied Volatilities $\bar{\sigma}_{N}^{SABR}(x)$, ($x=\frac{K-F}{F}$). The Put and Call price follows the Bachelier price with adjusted Volatility. For the Put price we get:
\begin{eqnarray}
P_{SABR}^{Hagan}(F, K, \bar{\sigma}_{1}) &\cong&F\times\\
& &\hspace{-2.5cm}   \left( x \mathrm{N}\left[ \frac{x}{\bar{\sigma}_{N}^{SABR}(x)} \right] + \frac{\bar{\sigma}_{N}^{SABR}(x)}{\sqrt{2\pi}}e^{-\frac{1}{2}\left( \frac{x}{\bar{\sigma}_{N}^{SABR}(x)} \right)^{2}} \right)\nonumber
\end{eqnarray}

The Cumulative probability has no known analytic expressions, but we easily use:

\begin{eqnarray}
D_{SABR}^{Hagan}(K) &=& \frac{\partial}{\partial K} P_{SABR}^{Hagan}(F, K, \bar{\sigma}_{1})\\
& =&\mathrm{N}\left[\frac{x}{\bar{\sigma}_{N}^{SABR}}\right]\\
& & + \left( \frac{\partial}{\partial K}\bar{\sigma}_{N}^{SABR}(x) \right)\frac{e^{-\frac{1}{2}\left( \frac{x}{\bar{\sigma}_{N}^{SABR}(x)} \right)^{2}}}{\sqrt{2\pi}}\nonumber
\end{eqnarray}

In our analysis we use the Hagan prices as an interpolation tool hence we do not adjust the parameters of the solutions to the parameters of the SABR stochastic equation, but we use parameters directly linked to observed prices:

\begin{eqnarray}
\bar{\sigma}_{N}^{SABR} (K) &=& \bar{\sigma}_{1}\frac{\xi}{H(\xi)} \\
\xi &=& \frac{\nu\sqrt{\Delta t}}{\bar{\sigma}_{1}} \frac{1}{1-\beta} \left(1 - (1-x)^{1-\beta}\right) \\
H(\xi) &=& \ln \left( \frac{\sqrt{1-2\xi\rho + \rho^{2}} + \xi - \rho}{1-\rho}\right)
\end{eqnarray}

For $\beta=0$ we get $\xi=x\nu\sqrt{\Delta T} / \bar{\sigma}_{1}$ while for $\beta=1$   we get $\xi = \ln(1+x)\nu\sqrt{\Delta T} / \bar{\sigma}_{1}$.

The SABR parametrization involves 4 parameters ($\bar{\sigma}_{1}$, $\rho$, $\nu$, $\beta$) with $-1 \leq \rho \leq 1$ and $0 \leq \beta \leq 1$. Again the first parameter is proportional to the ATM price, while the 3 others are describing the shape of the option price curve. Quite often we involve alternatively $\beta=0$ or $\beta=1$ , yet in our approach we treat all parameters as interpolation tools.

\item [(C)] Extrapolated SABR/SLN-HEX pricing

The SABR pricing is often efficient, yet still adjustments may be needed to improve modelling practical market prices.  We use a further extrapolation which is based on so-called HEX prices introduced some time ago by one of the author of this work \cite{JRK_cours}. The HEX (from Hyperbolic Exponential) option prices follows from the observation, that for any set of call and put option prices we can use the following expression for the prices:

\begin{eqnarray}
Call_{HEX} &=& \frac{\Pi_{ATM}}{\ln(2)}\left[1 + \exp(-\chi_{C}(x)\right)] \\
Put_{HEX} &=& \frac{\Pi_{ATM}}{\ln(2)}\left[1 + \exp(\chi_{P}(x)\right)]
\end{eqnarray}

If we use a linear expression for $\chi_{CP}(x) = -xF\ln(2)/\Pi_{ATM}$, we get a cumulative probability distribution being hyperbolic tangent.

The extension to the previously discussed SABR pricing goes via the formulas:

\begin{eqnarray}
\chi_{P}(x<0) &=& \chi_{P;\bar{\sigma}_{1},\rho,\nu,\beta}^{SABR}(x) + \vartheta^{3}_{L}(x)\exp(-a/x^{2})\nonumber \\ \\
\chi_{C}(x>0) &=& \chi_{C;\bar{\sigma}_{1},\rho,\nu,\beta}^{SABR}(x) + \vartheta^{3}_{R}(x)\exp(-a/x^{2})\nonumber\\
\end{eqnarray}
with $\vartheta^{3}_{L}(x)$ and $\vartheta^{3}_{R}(x)$ being two independent third order polynomial with no constant terms. The leading term $\chi_{\bar{\sigma}_{1},\rho,\nu,\beta}^{SABR}(x)$ is calculated from the SABR prices following:

\begin{eqnarray}
\chi_{P;\bar{\sigma}_{1},\rho,\nu,\beta}^{SABR}(x) + \ln \left( 2^{P_{SABR}^{Hagan}/\Pi_{ATM}}) -1 \right) \\
\chi_{C;\bar{\sigma}_{1},\rho,\nu,\beta}^{SABR}(x) + \ln \left( 2^{C_{SABR}^{Hagan}/\Pi_{ATM}}) -1 \right) 
\end{eqnarray}

We first establish the Hagan parameters  ($\bar{\sigma}_{1}$, $\rho$, $\nu$, $\beta$) and next improve the interpolation with adding the six parameters of $\vartheta^{3}_{L}(x)$ and $\vartheta^{3}_{R}(x)$.

We can also use the SABR/Hagan parameters corresponding to the SLN model (the SLN expressions are the only known exact solutions if the SABR stochastic equations) and add corrections following the HEX polynomials discussed above. Such a SLN-HEX extrapolation procedure proves often efficient and even better that the pure SABR interpolation.

\end{itemize}

To analyse market features related to pricing a realistic Structured Product, we consider a European S\&P500 Bonus Certificate priced 817 days ahead of expiry. This product has a shutting down barrier at 79.5\% of the initial underlying level.
On that day the structure of European Options on the S\&P500 nearly matching the maturity of the product had the following structure as compared to the SABR-HEX price interpolation FIG. \ref{fit_HEX_SABR}:

\begin{figure}[h!]
\centering
\includegraphics[width=.85\columnwidth,trim= 0mm 18mm 0mm 0mm ]{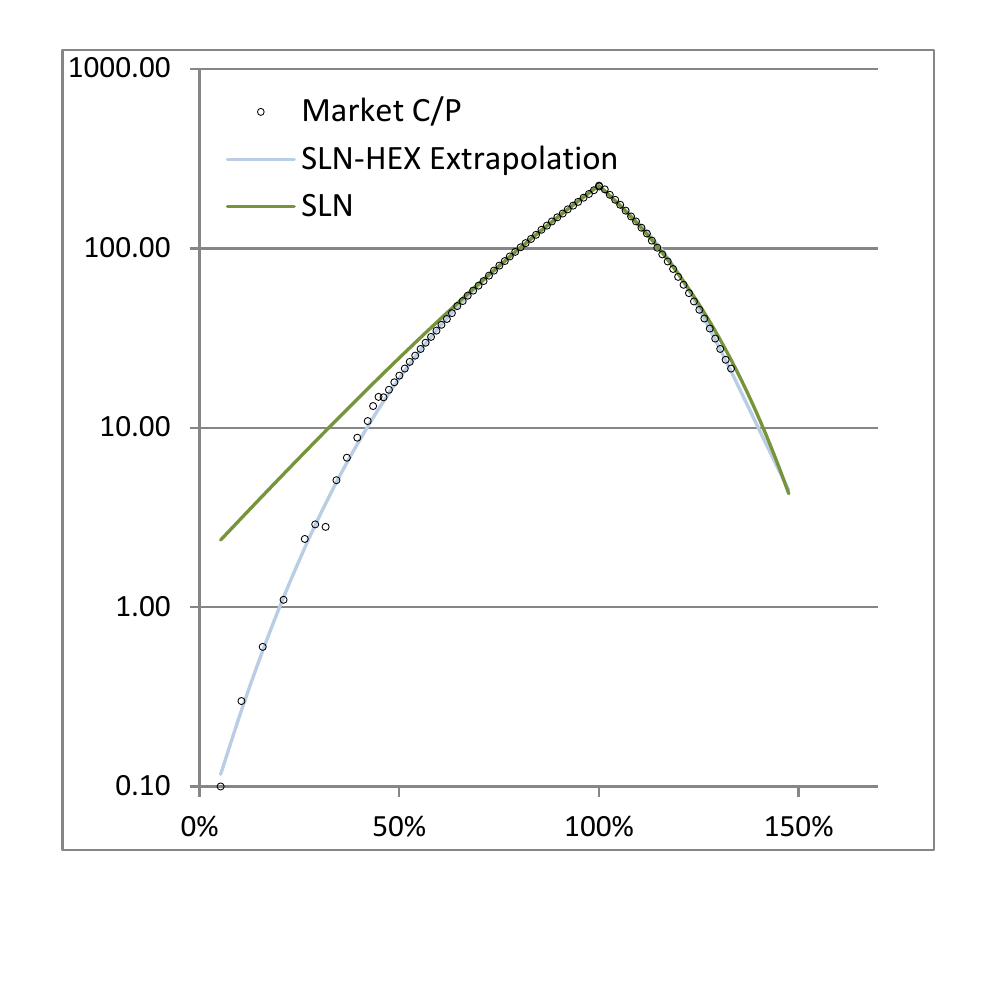}
\caption{\label{fit_HEX_SABR} Fitting results of the SLN model and HEX extrapolation corrections}
\end{figure}

As we see the SLN pricing gets correctly the near-ATM characteristics of the Volatility Surface, while the HEX extrapolation corrections allow getting a reasonable interpolation of far strikes. Eventually we deal with a 2+6 parameters interpolation which proves a bit more efficient that the 4 parameter SABR interpolation.
Pricing Structured products is dependent on the Put price interpolated at the Barrier, the Call Price interpolated at the Strike and the Cumulative probability of ending below the barrier ($p_{H-}$). The latter is proportional to the derivative of the interpolated Put price:

\begin{equation}
D_{model}^{interpolated}(K=B) = \frac{1}{DV}\frac{\partial}{\partial K}D_{model}^{interpolated}(K)
\end{equation}

Eventually, we need to interpolate very accurately the Put prices for $K \cong B$ and the Call prices for $K \cong S_{0} + R$. The comparison between the interpolation done using the SLN and SABR models looks as follows FIG. \ref{inter_SLN_SABR}:

\begin{figure}[h!]
\centering
\includegraphics[width=.85\columnwidth,trim= 0mm 18mm 0mm 0mm ]{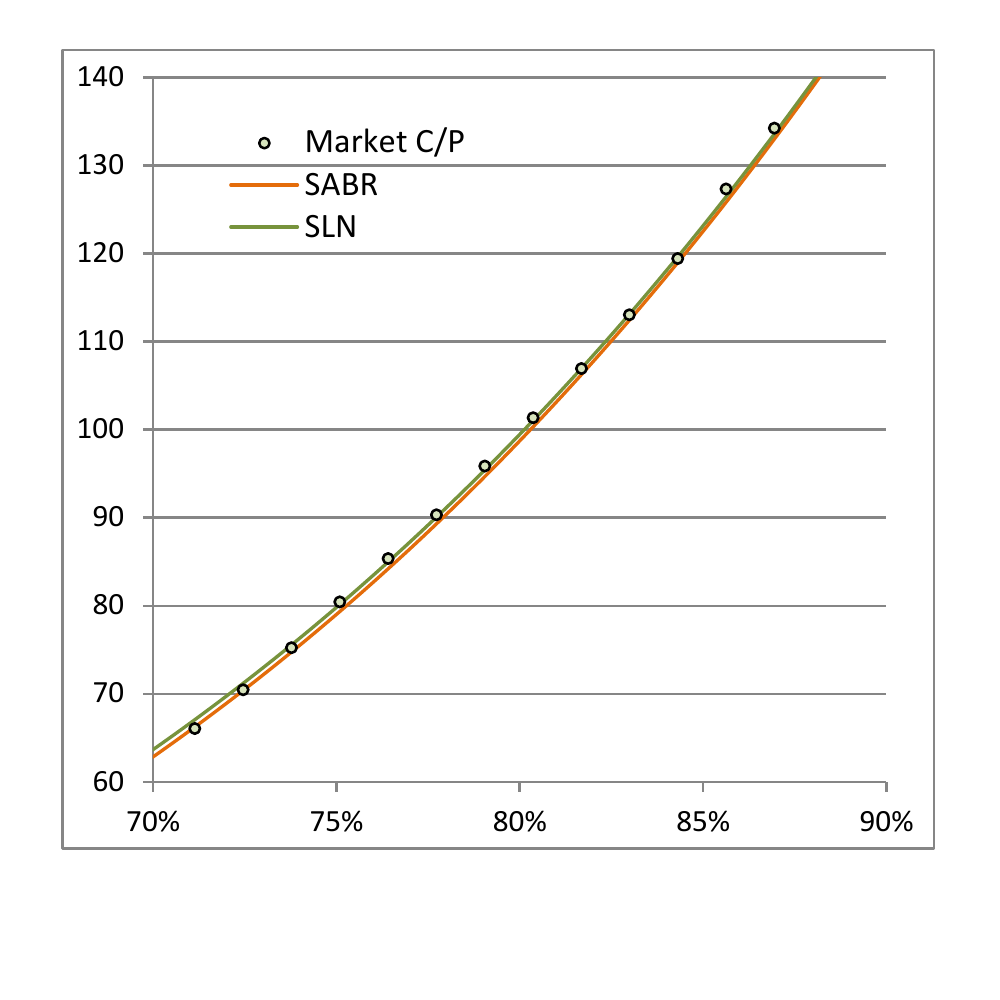}
\caption{\label{inter_SLN_SABR} SLN and SABR interpolation.}
\end{figure}

The SLN interpolation compared to the SLN-HEX interpolation looks as follows FIG. \ref{inter_SLN_HEX}:

\begin{figure}[h!]
\centering
\includegraphics[width=.85\columnwidth,trim= 0mm 10mm 0mm 0mm ]{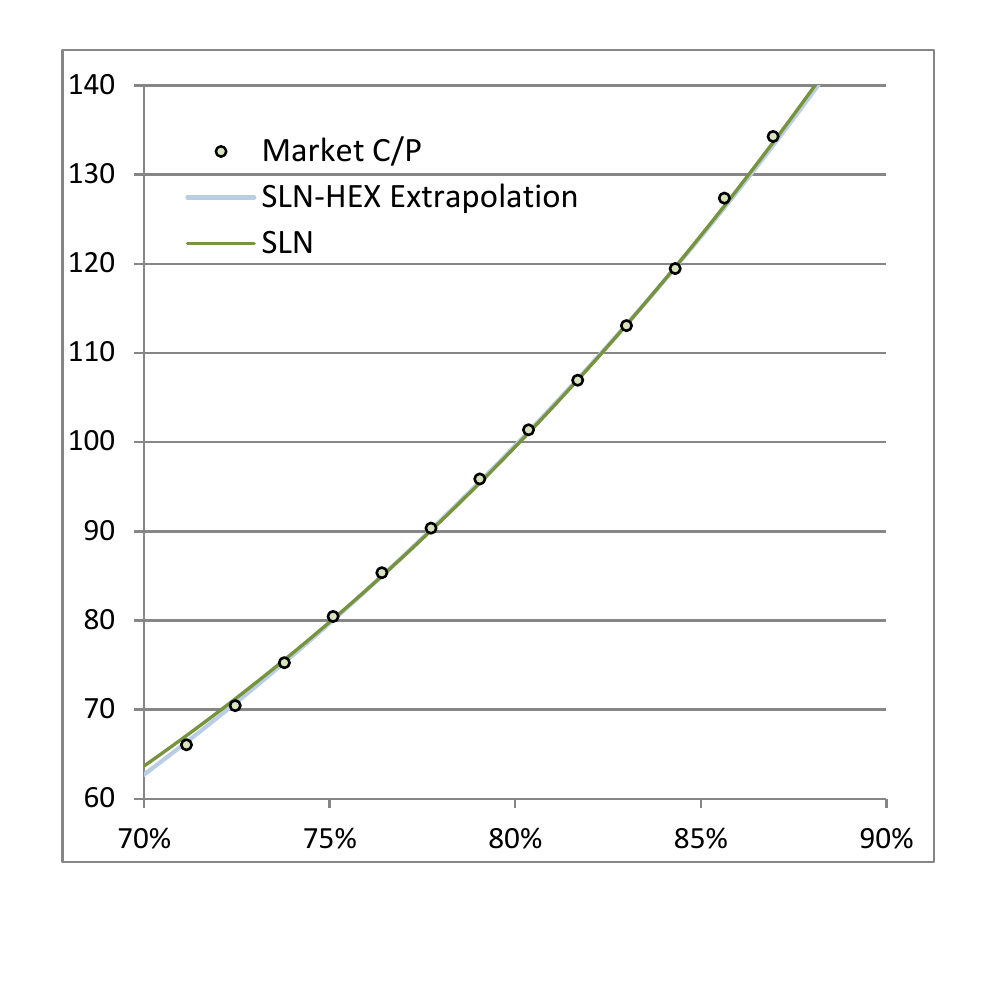}
\caption{\label{inter_SLN_HEX} SLN and SLN-HEX interpolation.}
\end{figure}

The end conclusion is that for pricing Vanilla Structured products, such as the Bonus Certificates and Barrier Reverse Convertible contracts discussed herein, using SLN near-ATM price interpolation to get an overall price accuracy not worse than $\pm1\%$. This follows the barriers lying in the range $0.65S_{0} \leq B \leq 0.8 S_{0}$ with volatility levels not higher than $\sigma_{1 \text{year}}^{ATM}\sqrt{\Delta T} \leq 30  \%$.

\section{Conditional probabilities and key approximations}
\label{appendixB}
Path dependent calculations require using martingale stochastic model. We typically use the following type of models ($F_{t, T}$ is the forward price curve):

\begin{eqnarray}
&\tilde{S}_{t} = F_{t,T}\, \tilde{X}_{t} \\
&d\tilde{X}_{t} = \tilde{\Lambda}(t, \tilde{\sigma}_{t}, q, \tilde{X}_{t})\,(\rho d\tilde{V}_{t} + \sqrt{1-\rho^{2}}d\tilde{W}_{t})
\end{eqnarray}

Where $\tilde{\Lambda}$ is dependent on the stochastic model. For the Shifted Log Normal model we have $\tilde{\Lambda}_{SLN} = \sigma_{t}(1 + q\tilde{f}_{t}$. We eventually shall use two alternative models to fit the Volatility surface:

\begin{eqnarray}
&\tilde{\Lambda}_{SLN} = \sigma_{t}\,(1 -q + q_{t}\tilde{X}_{t}) \\
&\tilde{\Lambda}_{SABR} = \sigma_{t}\,(\tilde{X}_{t})^{\beta}\exp \big( \nu\tilde{V}_{t} -\frac{\nu^{2}}{2}t \big)
\end{eqnarray}

Note that for $\sigma_{t} = \sigma_{0}$ and $\rho=\pm 1$ the SABR $\beta=0$ model is coincident with the SLN model \cite{JRK_D}.

We first calibrate the models to match option pricing for selected pricing dates and on that basis calculate the probability densities for two conditional probability distributions. Density of trajectories that have not hit the barrier $\rho_{BN}^{model}(\phi)$ and density of trajectories that have hit the barrier $\rho_{BH}^{model}(\phi)$.

We notice that:
   
\begin{eqnarray}
\rho_{BN}^{model}(\phi \geq B) + \rho_{BH}^{model}(\phi \geq B) &=& \rho_{T}^{interpolated}(\phi \geq B)\nonumber\\ \\
\rho_{BH}^{model}(\phi \leq B) &=& \rho_{T}^{interpolated}(\phi \leq B)\nonumber\\
\end{eqnarray}

Further we define the probability of hitting the barrier and ending above the barrier
    
\begin{eqnarray}
\rho_{H+}^{model} = \int_{B}^{\infty} d\phi\,\rho_{BH}^{SABR}(\phi)
\end{eqnarray}

And the probability of hitting the barrier and ending below the barrier:
   
\begin{equation}
\rho_{H-} = \int_{-\infty}^{B} d\phi\,\rho_{BH}^{model}(\phi) = \int_{-\infty}^{B}d\phi\,\rho_{T}^{interpolated}(\phi\leq B) 
\end{equation}

We stress again, that $\rho_{T}^{interpolated}(\phi)$ follows the interpolated market option prices (\ref{rho_interp_call}) and (\ref{rho_interp_put}). These probabilities densities are not conditional on hitting the barrier and are solely dependent on interpolated market option prices - hence ``model-independent". As discussed in Chapter 3, within the range of considered market and product parameters, numerical Monte-Carlo calculation of $p_{H+}^{model}$ is the only necessary ingredient to price American Bonus Certificates and Barrier Reverse Convertible contracts.

\section{Alternative stochastic calculation of $\delta^{model}$}
\label{appendixC}

Following the derived formula for \textit{ABC} the price:

\begin{equation}
\Pi_{ABC}^{model} \cong K - (K-B)(2 + \delta^{model})p_{H-} + Call(K, S_{0})
\end{equation}

and the parameters encountered in the calibrations discussed in this paper, price sensitivity is not larger than $\Delta\Pi_{ABC}^{model}/\Pi_{ABC}^{model}<\delta^{model}/20$ (the same result applies to the price of the ABRC contract.

Separately we have found that calibrations performed using the \textit{SLN} model are giving accurate enough price matching in the ranges determined by the barrier conditions (see \ref{appendixA}). Yet when such a price fit is made, the Volatility Surface is not matching the outcome of a static \textit{SLN} model. Two questions are arising:

\begin{itemize}
\item To what extend we can successfully use a \textit{SLN} model with time dependent $q_{t}$?
\item What is the accuracy sensitivity if we replace $q_{t}$ with $q_{0}$ being some sort of an average?
\end{itemize}

To elaborate a partial answer to these two questions, we have studied a fit to a 1 year Volatility Surface of SPX options. The Volatility Surface was modelled using a time dependent $q_{t}$ stochastic dynamic following

\begin{eqnarray}
\tilde{S}_{t} &=& S_{0}\,(1+\tilde{f}_{t}) \\
d\tilde{f}_{t} &=& \sigma_{t}\,(1+q_{t}\tilde{f}_{t})\,d\tilde{W}_{t}
\end{eqnarray}

With the following dynamic dependences:

\begin{eqnarray}
\label{sln_C}
\sigma_{t} &=& \sigma_{A}\,t^{\alpha} \\
q_{t} &=& -q_{B}\,t^{\beta}
\end{eqnarray}

And the parameters $\sigma_{A}=15\%$, $\alpha=0.124$, $q_{B}=-3.1$ and $\beta=-0.67$.

We have solved via Monte-Carlo the equation (\ref{sln_C}) and obtained the following $\delta$ ratios for various barrier levels:

\begin{table}[h!]
\centering
\begin{tabular}{ccccc}
Counts & Counts & Counts below & Counts above & $\delta_{d}$ \\
\hline
	60\% & 41101 & 18981 & 22120 & 0.17 \\ 
	65\% & 59763 & 27536 & 32227 & 0.17 \\ 
	70\% & 87772 & 40138 & 47634 & 0.19 \\ 
	75\% & 129513 & 58833 & 70680 & 0.20 \\ 
	80\% & 192882 & 86156 & 106726 & 0.24 \\ 
	90\% & 435379 & 188083 & 247296 & 0.31 \\
\end{tabular}
\caption{Number of trajectories that have hit the Barrier for various Barrier levels with dynamic $q$ simulation. The total amount of trajectories is $N=10^6$.}
\end{table}

Then we performed another Monte-Carlo simulation using a static $q_{0}=-3.82$ which corresponds to $q_{t=0.75}$ form the previous simulation. We obtained the following $\delta$ hit ratios:

\begin{table}[h!]
\centering
\begin{tabular}{ccccc}
Counts & Counts & Counts below & Counts above & $\delta_{s}$ \\
\hline
60\% & 47145 & 11248 & 13690 & 0.22\\ 
65\% & 65372 & 16296 & 19984 & 0.23\\ 
70\% & 91829 & 24227 & 29891 & 0.23\\ 
75\% & 131014 & 36727 & 45354 & 0.23\\ 
80\% & 190296 & 56764 & 71269 & 0.26\\ 
90\% & 422157 & 144681 & 193829 & 0.34\\
\end{tabular}
\caption{Number of trajectories that have hit the Barrier for various Barrier levels with static $q$ simulation. The total amount of trajectories is $N=10^6$.}
\end{table}

As we see, the maximal discrepancy of $\delta$ is equal to $\delta_{s}-\delta_{d} = \Delta\delta\leq 0.05$. This transaltes to a price sensitivity of:

\begin{eqnarray}
\frac{\Delta\Pi_{X}^{model}}{\Pi_{X}^{model}} < 0.4\%
\end{eqnarray}

On that basis we estimate, that the price sensitivity of the \textit{ABC}/\textit{ABRC} products is not importantly affected if we use a static \textit{q} model fitted to approximately $dT = \frac{3}{4}\Delta T$ of the product maturity. This is a general indication that the \textit{SLN} model is a very efficient pricing tool.

As a separate interesting conclusion, we found that for a range of parameters, the probability distributions of the time dependent $q_{t}$ model are very close to the one of the static Shifted Log-Normal model.

\clearpage

\section{Market Data- Graphs}
\label{appendixD}

\begin{table}
\begin{center}
\scalebox{0.8}{
\begin{tabular}{ |c|c|c|c|c| } 
\hline
Price Date & Maturity Date & Time to maturity(days) & \# Strikes \\
\hline
\multirow{10}{8em}{02/02/2015} & 06/03/2015 & 32 & 171\\
& 10/04/2015 & 67 &	164\\
& 15/05/2015 & 102 & 100\\
& 30/06/2015 & 148 & 60\\
& 18/09/2015 & 228 & 74\\
& 18/12/2015 & 319 &	95\\
& 15/01/2016 & 347 &	75\\
& 17/06/2016 & 501 &	81\\
& 16/12/2016 & 683 &	84\\
& 15/12/2017 & 1047 & 92\\
\hline
\multirow{10}{8em}{24/03/2015} & 24/04/2015	& 31 &	177\\
& 29/05/2015 & 66 &	179\\
& 30/06/2015 &	98	& 59\\
& 31/07/2015 &	129	& 60\\
& 18/09/2015 &	178	& 68\\
& 15/01/2016 &	297	& 75\\
& 18/03/2016 &	360	& 80\\
& 17/06/2016 &	451	& 79\\
& 16/12/2016 &	633	& 82\\
& 15/12/2017 &	997	& 90\\
\hline
\multirow{10}{8em}{26/05/2015} & 26/06/2015	& 31 & 177\\
& 31/07/2015	&	66	&	173\\
& 28/08/2015	&	94	&	180\\
& 30/09/2015	&	127	&	63\\
& 15/01/2016	&	234	&	75\\
& 18/03/2016	&	297	&	81\\
& 31/03/2016	&	310	&	36\\
& 17/06/2016	&	388	&	82\\
& 16/12/2016	&	570	&	84\\
& 15/12/2017	&	934	&	92\\
\hline
\multirow{10}{8em}{24/08/2015} & 30/09/2015	& 37 & 180\\
& 30/10/2015	&	67	&	182\\
& 27/11/2015	&	95	&	177\\
& 15/01/2016	&	144	&	68\\
& 18/03/2016	&	207	&	79\\
& 17/06/2016	&	298	&	76\\
& 12/06/2016	&	480	&	81\\
& 20/01/2017	&	515	&	95\\
& 16/06/2017	&	662	&	89\\
& 15/12/2017	&	844	&	90\\
\hline
\multirow{2}{8em}{20/09/2015} & 16/06/2017	& 635 & 90\\
& 15/12/2017 & 817 & 92\\
\hline
\multirow{2}{8em}{22/10/2015} & 16/06/2017	& 603 & 88\\
& 15/12/2017 & 785 & 92\\
\hline
\multirow{2}{8em}{30/11/2015} & 16/06/2017	& 564 & 87\\
& 15/12/2017 & 746 & 91\\
\hline
\end{tabular}
}
\end{center}
\caption{S\&P 500 options data set}
\label{table:1}
\end{table}

The above table is a presentation of the data set of S\&P500 calls and puts that were used to price the European Bonus Certificate. The reason why we chose to have at least 10 maturities for the first 4 dates has to do with the fact that the product matures on 24/07/2017 and as can be seen, for the first 3 pricing dates the only maturities close to that date are 16/12/2017 and 15/12/2017 which is quite far from the product maturity. Hence, a bigger set of maturities is required in order to obtain more accurate interpolation results. For the last 3 pricing dates (also for the 4th one) we could find options data maturing on 16/06/2017 which is very close to the product maturity. Hence, it is harmless to assume that the accuracy of the interpolated parameters is high in that case.

\begin{table}
\begin{center}
\scalebox{0.8}{
\begin{tabular}{ |c|c|c|c|c| } 
\hline
Price Date & Maturity Date & Time to maturity(days) & \# Strikes \\
\hline
\multirow{7}{8em}{08/07/2014} 
& 15/08/2014 & 38 & 21\\
& 19/09/2014	&	73	&	30\\
& 19/12/2014	&	164	&	54\\
& 20/03/2015	&	255	&	21\\
& 18/12/2015	&	528	&	71\\
& 17/06/2015	&	710	&	49\\
& 16/12/2015	&	892	&	70\\
\hline
\multirow{7}{8em}{08/08/2014} 
& 19/09/2014	&	42	&	26\\
& 17/10/2014	&	70	&	22\\
& 19/12/2014	&	133	&	67\\
& 20/03/2015	&	224	&	21\\
& 19/06/2015	&	315	&	23\\
& 18/12/2015	&	497	&	68\\
& 16/12/2016	&	861	&	70\\
\hline
\multirow{7}{8em}{19/09/2014}
& 17/10/2014	&	28	&	21\\
& 21/11/2014	&	63	&	24\\
& 19/12/2014	&	91	&	60\\
& 20/03/2015	&	182	&	21\\
& 19/06/2015	&	273	&	24\\
& 18/12/2015	&	455	&	66\\
& 17/06/2016	&	637	&	51\\
& 16/12/2016	&	819	&	70\\
\hline
\multirow{7}{8em}{28/11/2014}
& 16/01/2015	&	49	&	33\\
& 20/02/2015	&	84	&	27\\
& 20/03/2015	&	112	&	46\\
& 19/06/2015	&	203	&	27\\
& 18/09/2015	&	294	&	29\\
& 18/12/2015	&	385	&	67\\
& 16/12/2016	&	749	&	68\\
\hline
\multirow{7}{8em}{03/02/2015}
& 20/03/2015	&	45	&	27\\
& 17/04/2015	&	73	&	33\\
& 19/06/2015	&	136	&	45\\
& 18/09/2015	&	227	&	32\\
& 18/12/2015	&	318	&	69\\
& 17/06/2016	&	500	&	40\\
& 16/12/2016	&	682	&	68\\
& 16/06/2017	&	864	&	56\\
\hline
\multirow{7}{8em}{01/06/2015}
& 17/07/2015	&	46	&	43\\
& 21/08/2015	&	81	&	49\\
& 18/12/2015	&	200	&	51\\
& 18/03/2016	&	291	&	44\\
& 17/06/2016	&	382	&	51\\
& 16/12/2016	&	564	&	76\\
& 16/06/2017	&	746	&	46\\
\hline
\multirow{7}{8em}{01/09/2015}
& 16/10/2015	&	45	&	59\\
& 20/11/2015	&	80	&	41\\
& 18/12/2015	&	108	&	68\\
& 17/06/2016	&	290	&	82\\
& 16/12/2016	&	472	&	76\\
& 16/06/2017	&	654	&	70\\
\hline
\end{tabular}
}
\end{center}
\caption{EUROSTOXX 50 options data set}
\label{table:5.2}
\end{table}


\begin{figure*}
\centering
  \begin{subfigure}{1\columnwidth}
  \includegraphics[width=1\columnwidth]{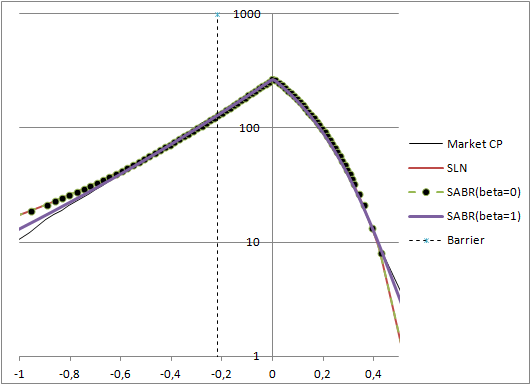}
  \caption{Fitting results on OTM options\label{fig:fit 1}}
  \end{subfigure}
  \hfill
  \begin{subfigure}{1\columnwidth}
  \includegraphics[width=1\columnwidth]{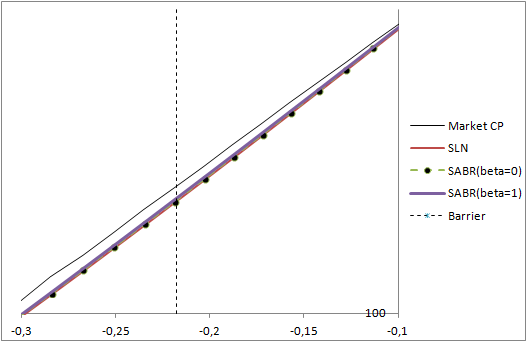}
  \caption{Zoomed fitted prices around the Barrier Level\label{fig:zoom 1}}
  \end{subfigure}
\caption{S\&P 500 Options on pricing date 02/02/2015 with time to maturity T=2.86 years}
\end{figure*}

\begin{table*}
\centering
\begin{tabular}{lc|cccccccccc}
\hline
\multicolumn{2}{c|}{\textbf{Price Date}}                   & \multicolumn{10}{c}{02/02/2015}                                                      \\ \hline
\multicolumn{2}{c|}{\textbf{Time to maturity(days)}}      & 32     & 67     & 102    & 148  & 228    & 319    & 347    & 501    & 683    & 1047   \\ \hline
SLN            & $\boldsymbol{q}$                & -11.01 & -7.7   & -6.07  & -4.7 & -3.16  & -2.7   & -2.57  & -1.82  & -1.08  & -0.5   \\
               & $\boldsymbol{\bar{\sigma}_{q}}$ & 0.0508 & 0.0744 & 0.0943 & 0.11 & 0.1464 & 0.1698 & 0.1768 & 0.2169 & 0.2642 & 0.3399 \\
   &    &   &   &  &  &  &   &   &  &   &   \\
SABR $\beta=1$ & $\boldsymbol{\rho}$             & -0.85  & -0.84  & -0.88  & -0.9 & -0.99  & -0.99  & -0.99  & -0.99  & -0.99  & -0.99  \\
               & $\boldsymbol{\nu}$              & 2.37   & 1.8    & 1.15   & 0.91 & 0.58   & 0.49   & 0.46   & 0.33   & 0.21   & 0.1   
\end{tabular}
\caption{Calibrated Parameters SLN and SABR($\beta=1$) \label{table:5.3}}
\end{table*}

\begin{figure*}
    \centering
    \includegraphics[width=1.01\columnwidth]{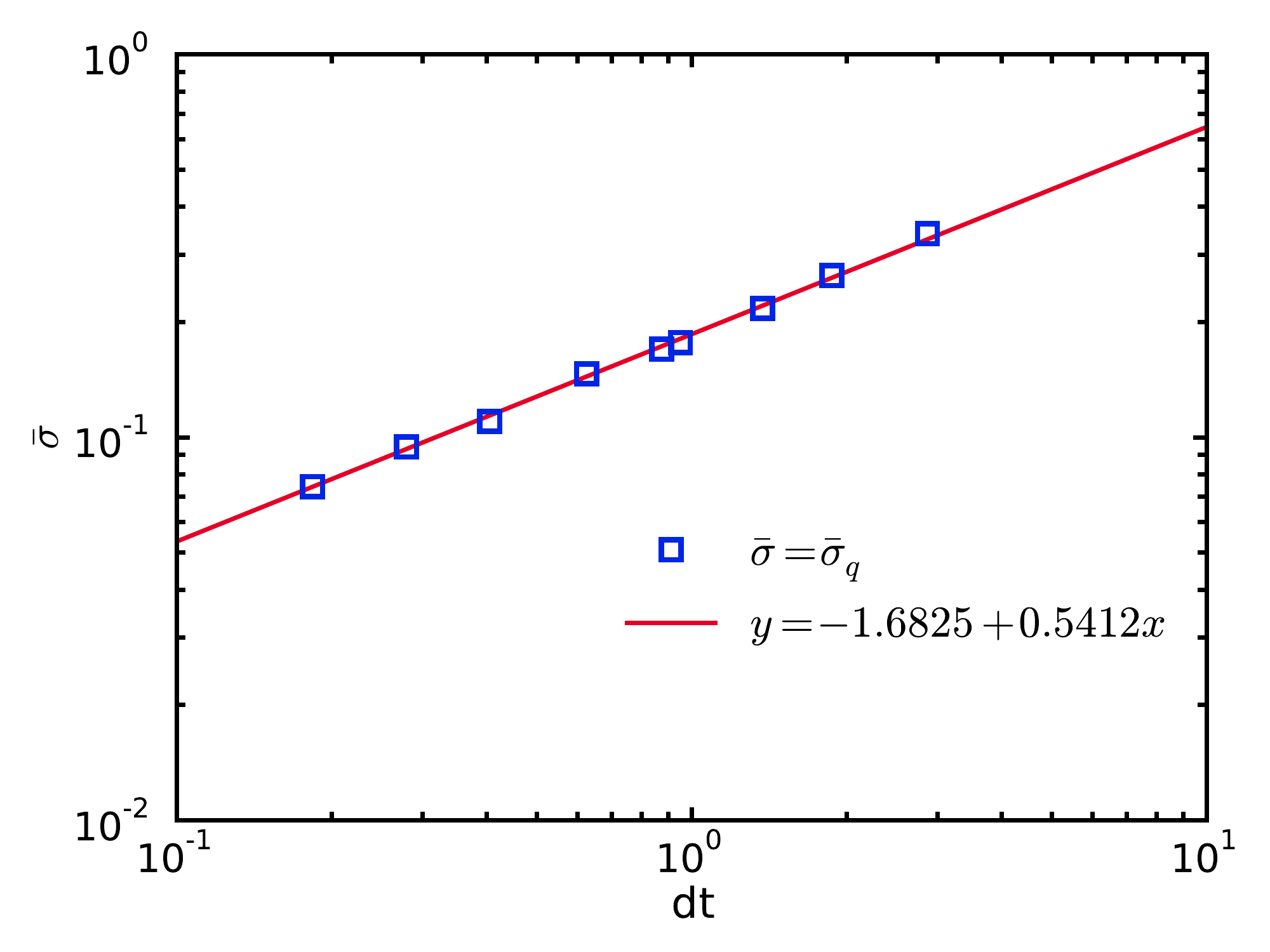}
    \includegraphics[width=1.01\columnwidth]{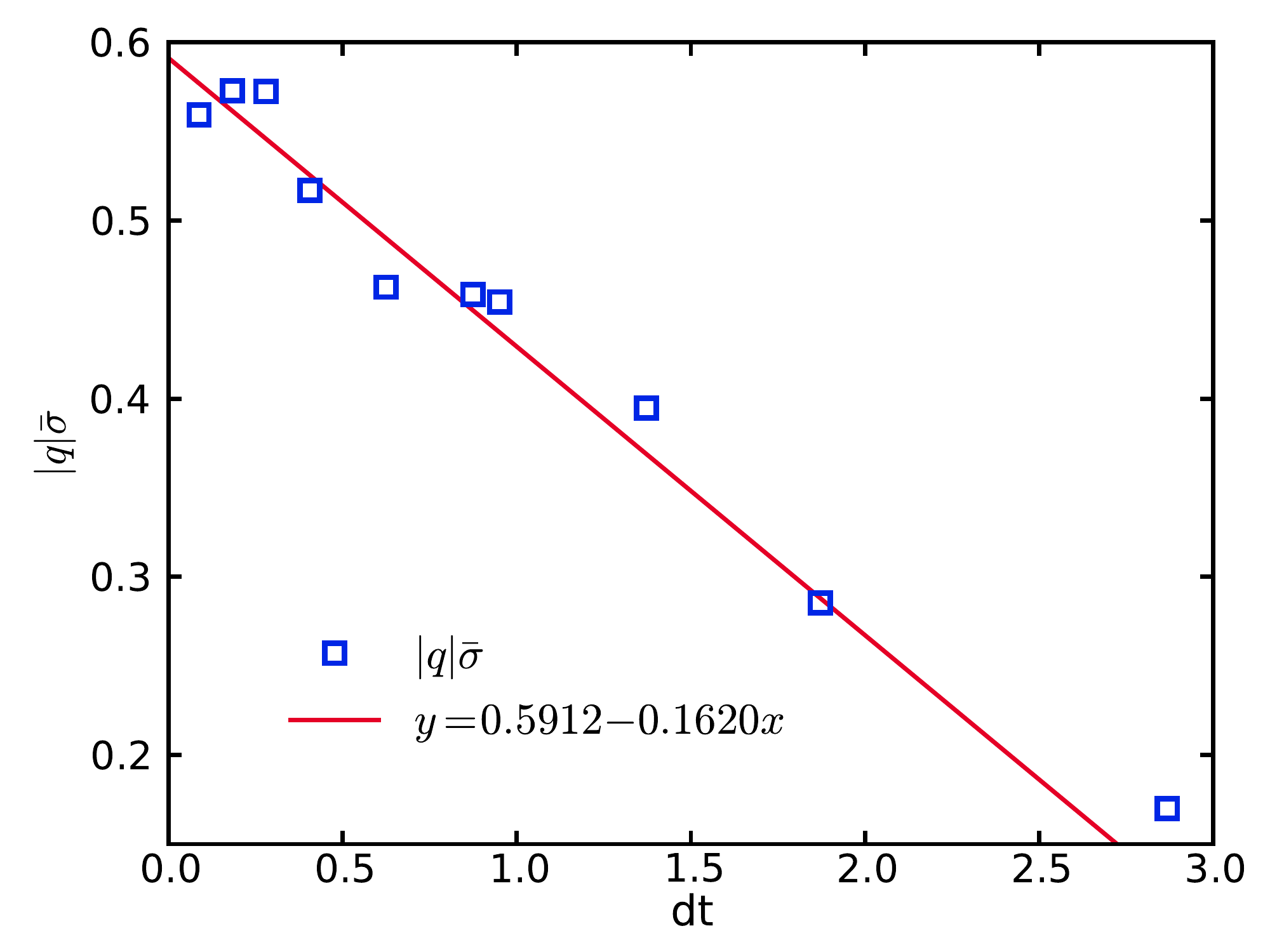}
    \caption{Log-log plot of $\bar{\sigma}=\bar{\sigma}_{q}$ (left), Lin-lin plot of  $|q|\bar{\sigma}_{q}$ (right). S\&P500, price date: 02/02/2015}
\end{figure*}


\begin{figure*}
  \begin{subfigure}{1\columnwidth}
  \centering
  \includegraphics[width=1\columnwidth]{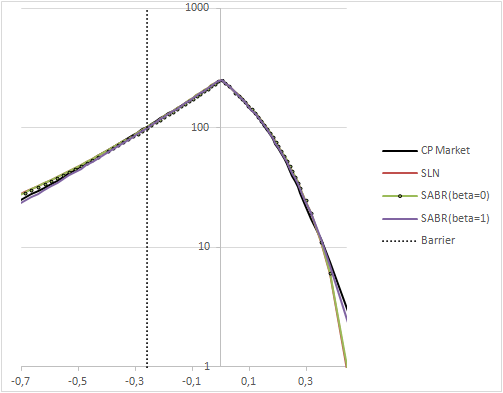}
  \caption{Fitting results on OTM options }
  \label{fig:fit 2}
  \end{subfigure}
  \hfill
  \begin{subfigure}{1\columnwidth}
  \centering
  \includegraphics[width=1\columnwidth]{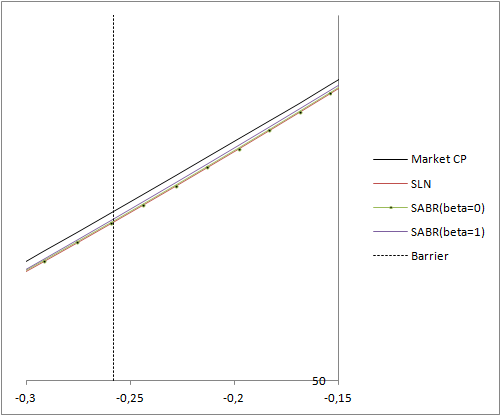}
  \caption{Zoomed fitted prices around the Barrier Level}
  \label{fig:zoom 2}
  \end{subfigure}
\caption{S\&P 500 Options on pricing date 24/03/2015 with time to maturity T=2.71 years}
\end{figure*}

\begin{table*}[H]
\begin{tabular}{lc|cccccccccc}
\cline{1-12}
\multicolumn{2}{c}{\textbf{Price Date}}             & \multicolumn{10}{|c}{24/03/2015}                                                        \\ \cline{1-12} 
\multicolumn{2}{c|}{\textbf{Time to maturity(days)}} & 31     & 66     & 98     & 129    & 178    & 297    & 360    & 451    & 633    & 997    \\ \cline{1-12} 
SLN               & $\boldsymbol{q}$                  & -18.13 & -9.99  & -8.36  & -5.61  & -4.12  & -3.24  & -2.36  & -2.31  & -1.47  & -0.75  \\
                  & $\boldsymbol{\bar{\sigma}_{q}}$   & 0.0335 & 0.0578 & 0.0738 & 0.0995 & 0.1161 & 0.1482 & 0.1682 & 0.1903 & 0.2302 & 0.3067 \\
   &    &   &   &  &  &  &   &   &  &   &   \\
SABR $\beta=1$    & $\boldsymbol{\rho}$               & -0.83  & -0.91  & -0.99  & -0.99  & -0.99  & -0.99  & -0.99  & -0.99  & -0.99  & -0.99  \\
                  & $\boldsymbol{\nu}$                & 2.30   & 1.43   & 1.17   & 1.09   & 0.67   & 0.52   & 0.44   & 0.39   & 0.25   & 0.007 
\end{tabular}
\caption{Calibrated Parameters SLN and SABR($\beta=1$)}
\label{table:5.4}
\end{table*}

\begin{figure*}
    \centering
    \includegraphics[width=1.01\columnwidth]{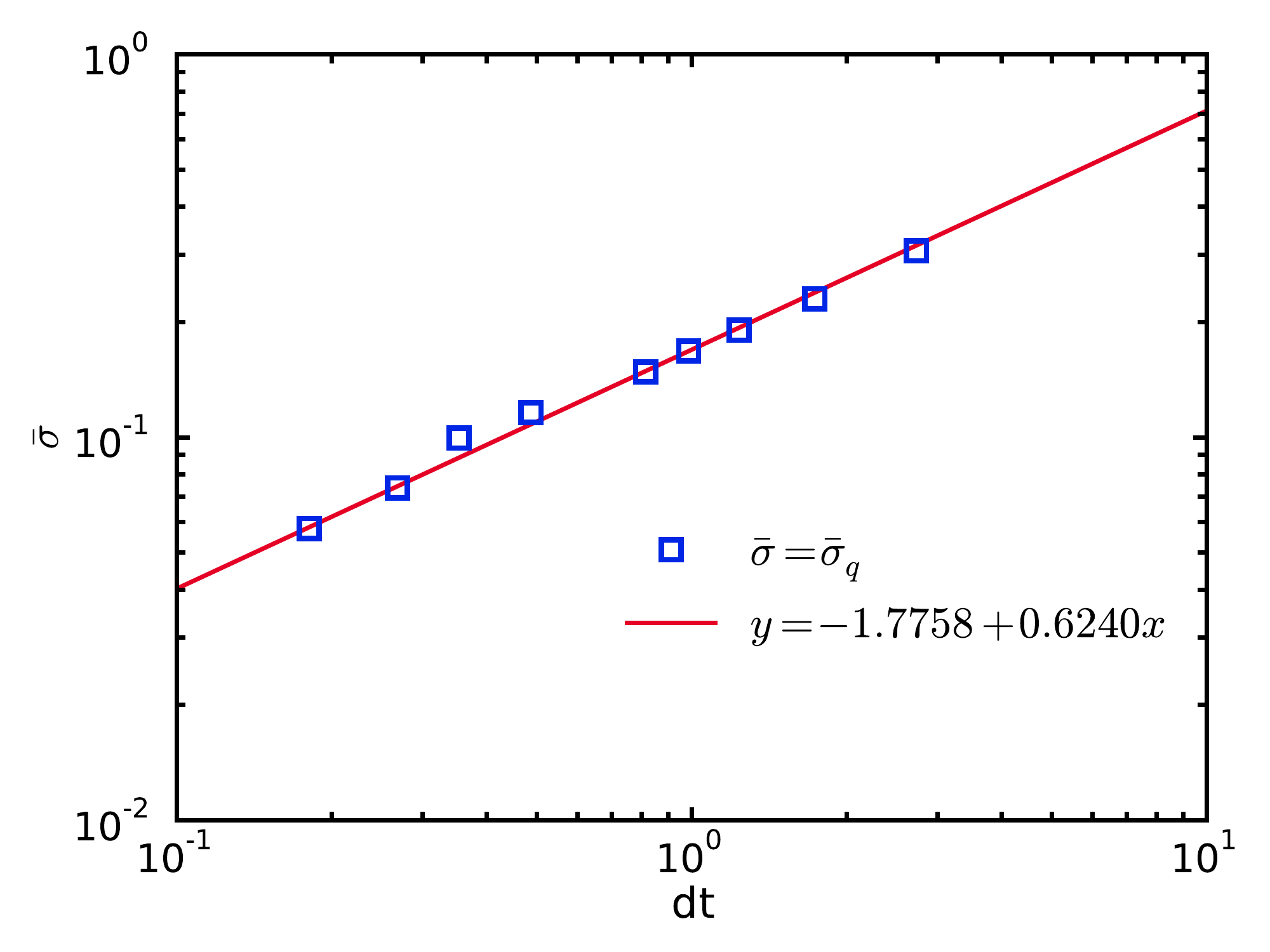}
    \includegraphics[width=1.01\columnwidth]{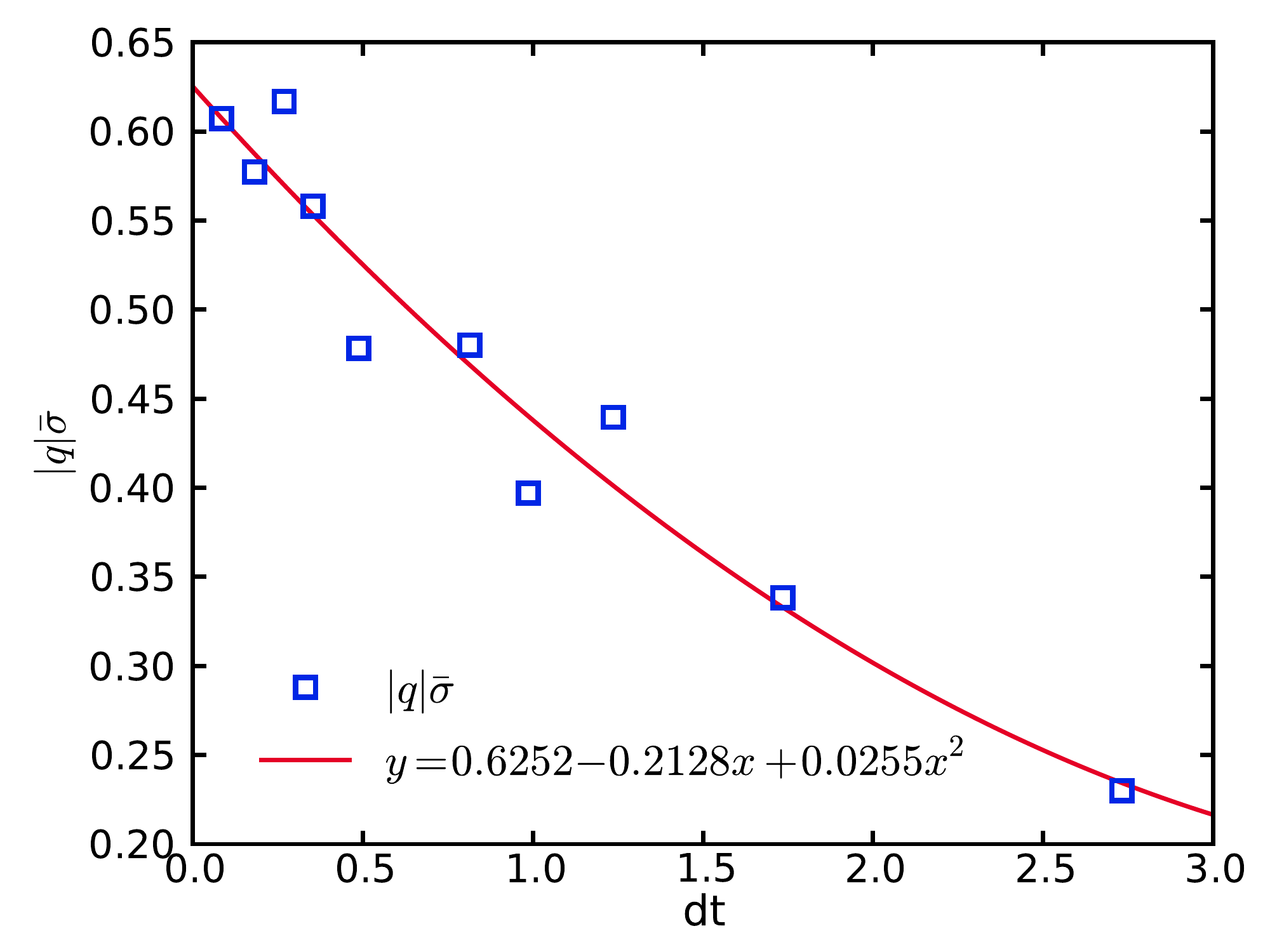}
    \caption{Log-log plot of $\bar{\sigma}=\bar{\sigma}_{q}$ (left), Lin-lin plot of  $|q|\bar{\sigma}_{q}$ (right). S\&P500, price date: 24/03/2015}
\end{figure*}

\begin{figure*}
\centering
\includegraphics[width=1.1\columnwidth]{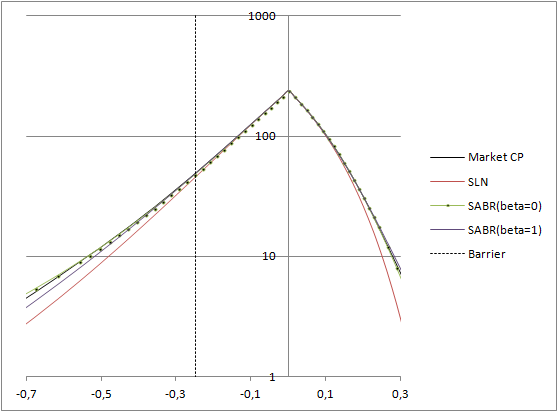} 
\caption{figure}{Fitting Results of EUROSTOXX 50 Index Options on pricing date 28/11/2014 with time to maturity T=1.05 years}
\end{figure*}

\begin{table*}[H]
\begin{tabular}{lc|cccccccc}
\cline{1-9}
\multicolumn{2}{c}{\textbf{Price Date}}             & \multicolumn{7}{|c}{28/11/2014}                                                        \\ \cline{1-9} 
\multicolumn{2}{c|}{\textbf{Time to maturity(days)}}      & 49     & 84     & 112    & 203    & 294    & 385    & 749     \\ \hline
SLN            & $\boldsymbol{q}$                & -4.23  & -2.92  & -2.78  & -1.73  & -1.1   & -0.73  & -0.0001 \\
               & $\boldsymbol{\bar{\sigma}_{q}}$ & 0.0573 & 0.0804 & 0.0952 & 0.1312 & 0.1619 & 0.1885 & 0.274   \\
               &                                 &        &        &        &        &        &        &         \\
SABR $\beta=1$ & $\boldsymbol{\rho}$             & -0.53  & -0.61  & -0.65  & -0.7   & -0.66  & -0.56  & -0.54   \\
               & $\boldsymbol{\nu}$              & 2.51   & 1.58   & 1.24   & 0.87   & 0.52   & 0.46   & 0.41   
\end{tabular}
\caption{Calibrated Parameters SLN and SABR($\beta=1$)}
\label{table:5.5}
\end{table*}

\begin{figure*}
    \centering
    \includegraphics[width=1.01\columnwidth]{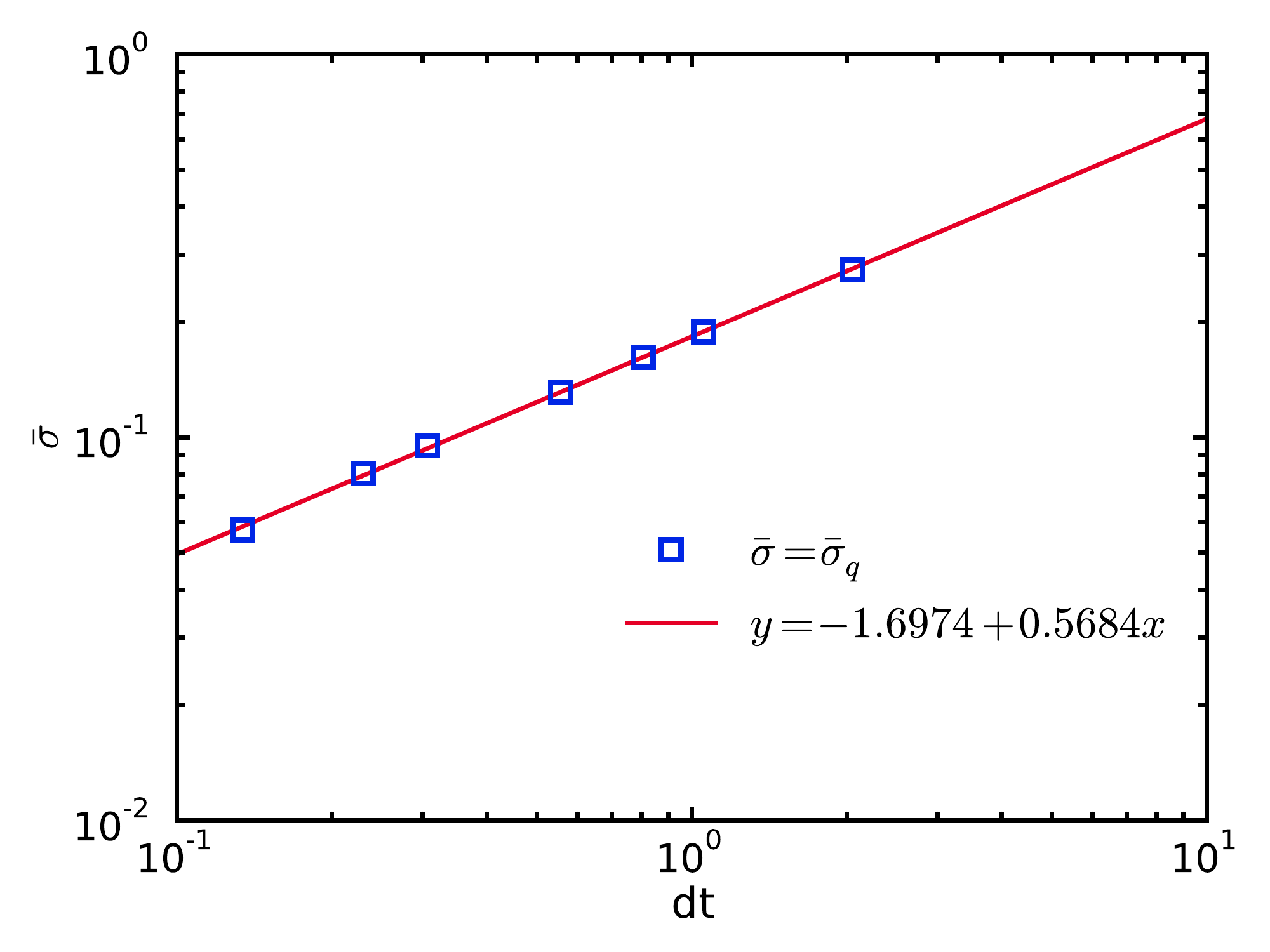}
    \includegraphics[width=1.01\columnwidth]{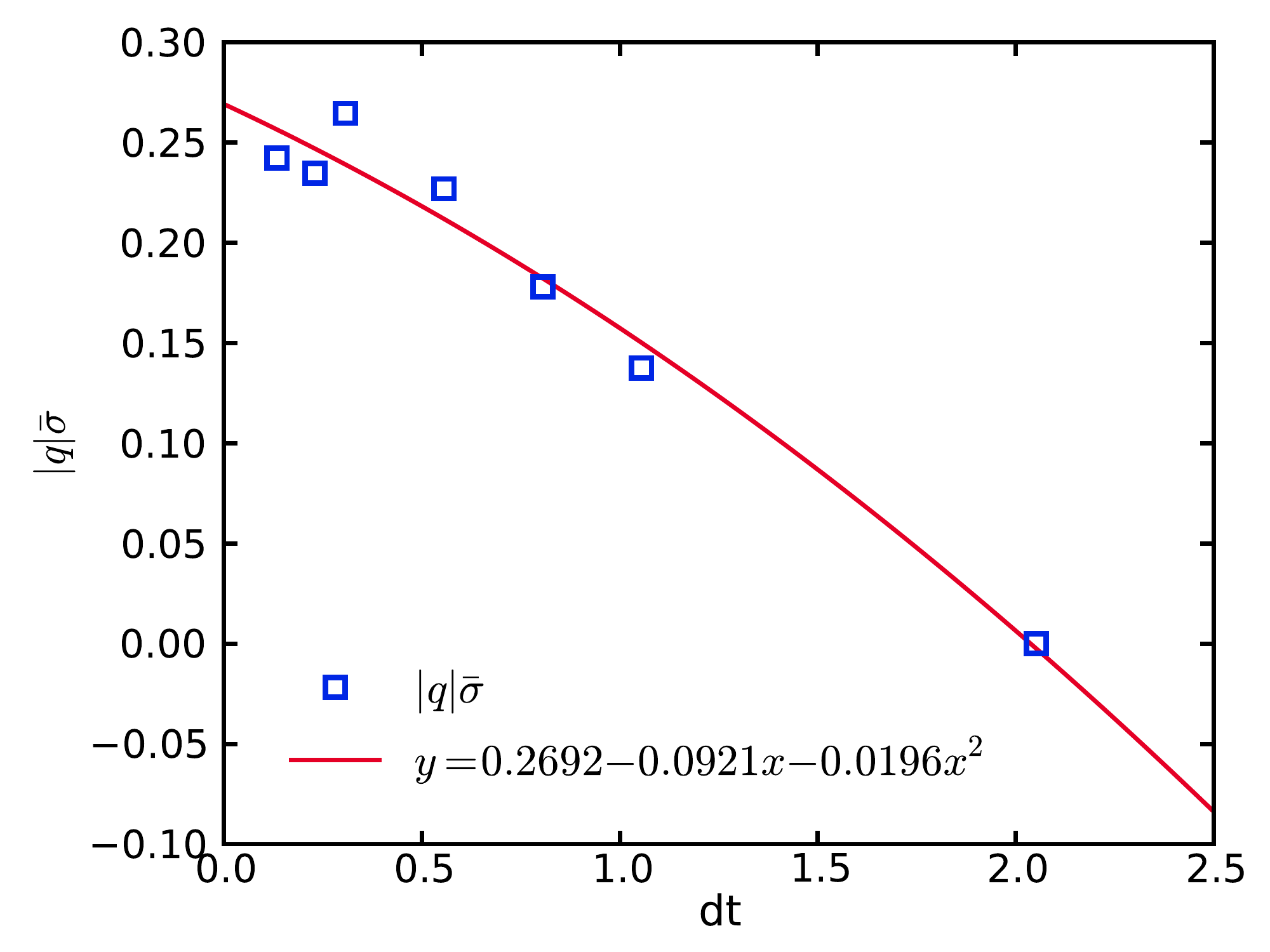}
    \caption{Log-log plot of $\bar{\sigma}=\bar{\sigma}_{q}$ (left), Lin-lin plot of  $|q|\bar{\sigma}_{q}$ (right). EUROSTOXX 50, price date: 28/11/2014}
\end{figure*}


\begin{figure*}
\centering
\includegraphics[width=1.1\columnwidth]{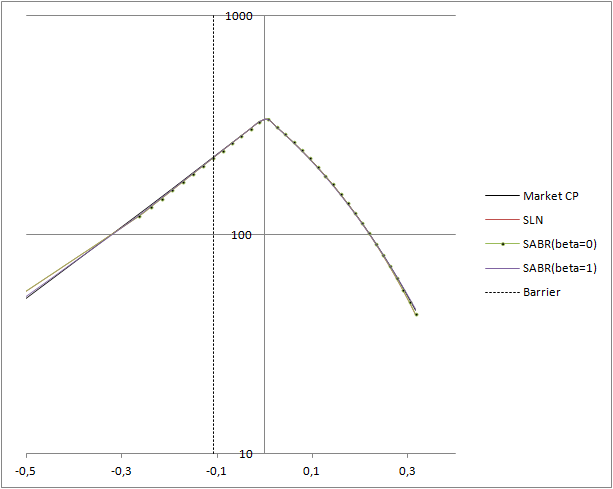} 
\caption{figure}{Fitting Results of EUROSTOXX 50 Index Options on pricing date 03/02/2015 with time to maturity T=1.86 years}
\end{figure*}

\begin{table*}[H]
\begin{tabular}{lc|cccccccc}
\cline{1-9}
\multicolumn{2}{c}{\textbf{Price Date}}             & \multicolumn{7}{|c}{03/02/2015}                                                        \\ \cline{1-9} 
\multicolumn{2}{c|}{\textbf{Time to maturity(days)}}      & 45     & 73     & 136    & 227    & 318    & 500    & 682        \\ \hline
SLN            & $\boldsymbol{q}$                & -4.84  & -3.63  & -2.4   & -1.54  & -0.36  & -0.22  & -0.09    \\
               & $\boldsymbol{\bar{\sigma}_{q}}$ & 0.0714 & 0.0913 & 0.1214 & 0.1654 & 0.1993 & 0.2134 & 0.2823  \\
               &                                  
                &  &  &  &  &        &        &        \\
SABR $\beta=1$ & $\boldsymbol{\rho}$             & -0.68  & -0.68  & -0.59  & -0.71  & -0.75  & -0.69  & -0.67   \\
               & $\boldsymbol{\nu}$              & 1.83   & 1.45   & 1.2    & 0.81   & 0.63   & 0.54   & 0.48     
\end{tabular}
\caption{Calibrated Parameters SLN and SABR($\beta=1$)}
\label{table:5.6}
\end{table*}

\begin{figure*}
    \centering
    \includegraphics[width=1.01\columnwidth]{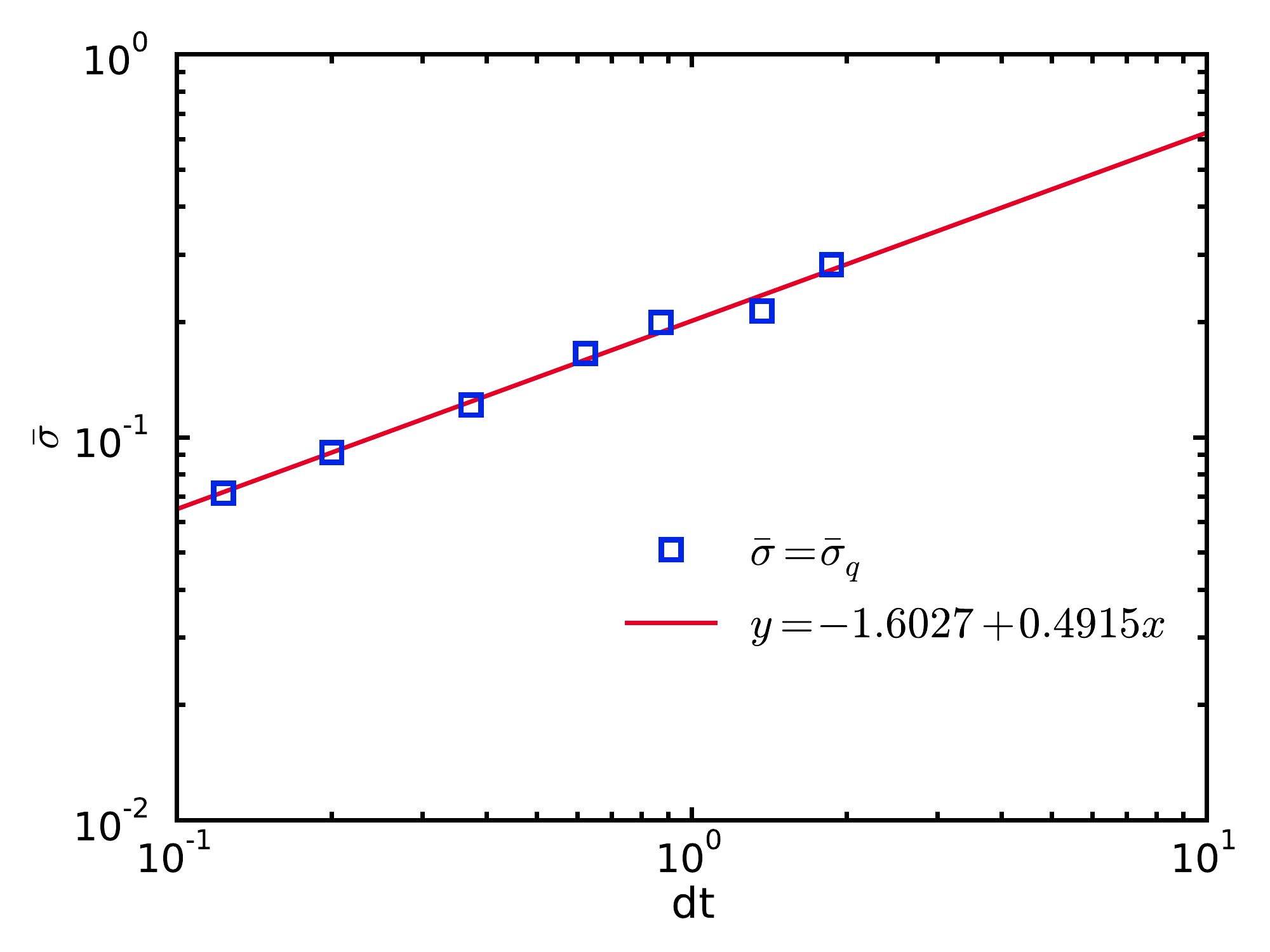}
    \includegraphics[width=1.01\columnwidth]{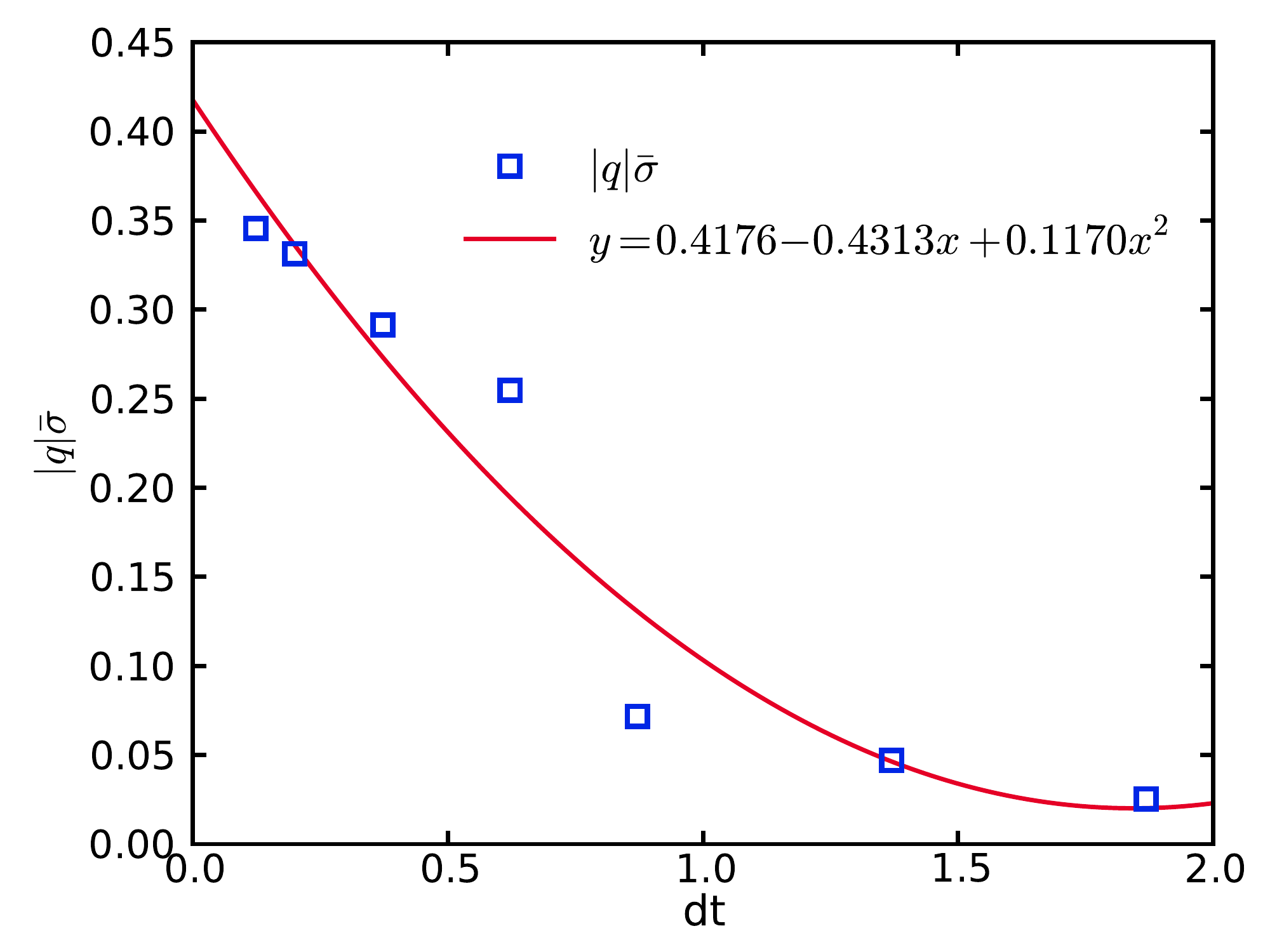}
    \caption{Log-log plot of $\bar{\sigma}=\bar{\sigma}_{q}$ (left), Lin-lin plot of  $|q|\bar{\sigma}_{q}$ (right). EUROSTOXX 50, price date: 03/02/2015}
\end{figure*}


\begin{figure*}
\centering
\includegraphics[width=1.1\columnwidth]{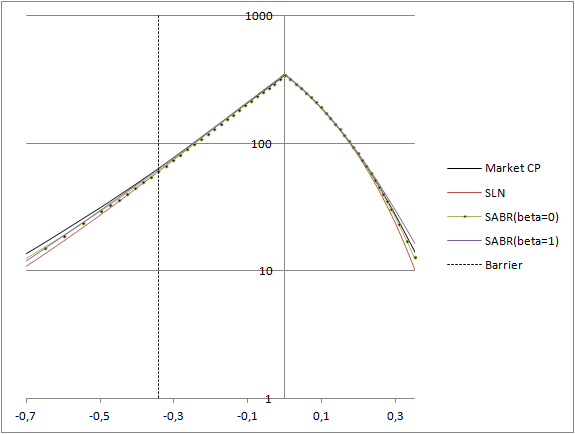} 
\caption{figure}{Fitting Results of EUROSTOXX 50 Index Options on pricing date 01/06/2015 with time to maturity T=1.54 years}
\end{figure*}

\begin{table*}[H]
\begin{tabular}{lc|ccccccc}
\cline{1-8}
\multicolumn{2}{c}{\textbf{Price Date}}             & \multicolumn{6}{|c}{01/06/2015}                                                        \\ \cline{1-8} 
\multicolumn{2}{c|}{\textbf{Time to maturity(days)}}      & 46     & 81     & 291    & 382    & 564    & 746        \\ \hline
SLN            & $\boldsymbol{q}$                & -3.08  & -2.61  & -1.05  & -0.59  & -0.34  & -0.0001  \\
               & $\boldsymbol{\bar{\sigma}_{q}}$ & 0.0786 & 0.0994 & 0.1791 & 0.2065 & 0.2469 & 0.2919     \\
               &                                 &        &        &        &        &        &         &         \\
SABR $\beta=1$ & $\boldsymbol{\rho}$             & -0.68  & -0.67  & -0.71  & -0.67  & -0.61  & -0.51     \\
               & $\boldsymbol{\nu}$              & 1.41   & 1.2    & 0.63   & 0.53   & 0.48   & 0.41       
\end{tabular}
\caption{Calibrated Parameters SLN and SABR($\beta=1$)}
\label{table:5.7}
\end{table*}
       
\begin{figure*}
    \centering
    \includegraphics[width=1.01\columnwidth]{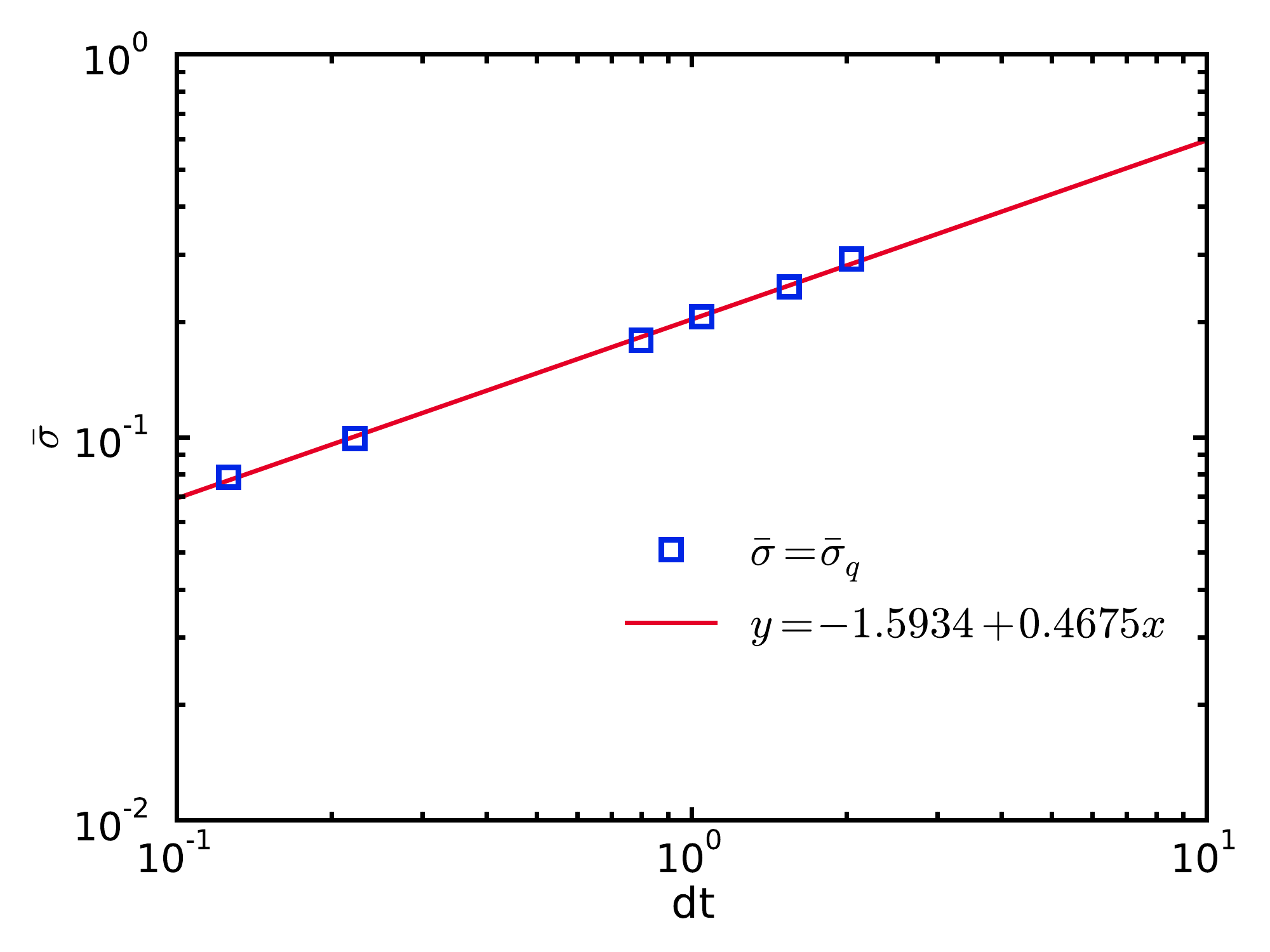}
    \includegraphics[width=1.01\columnwidth]{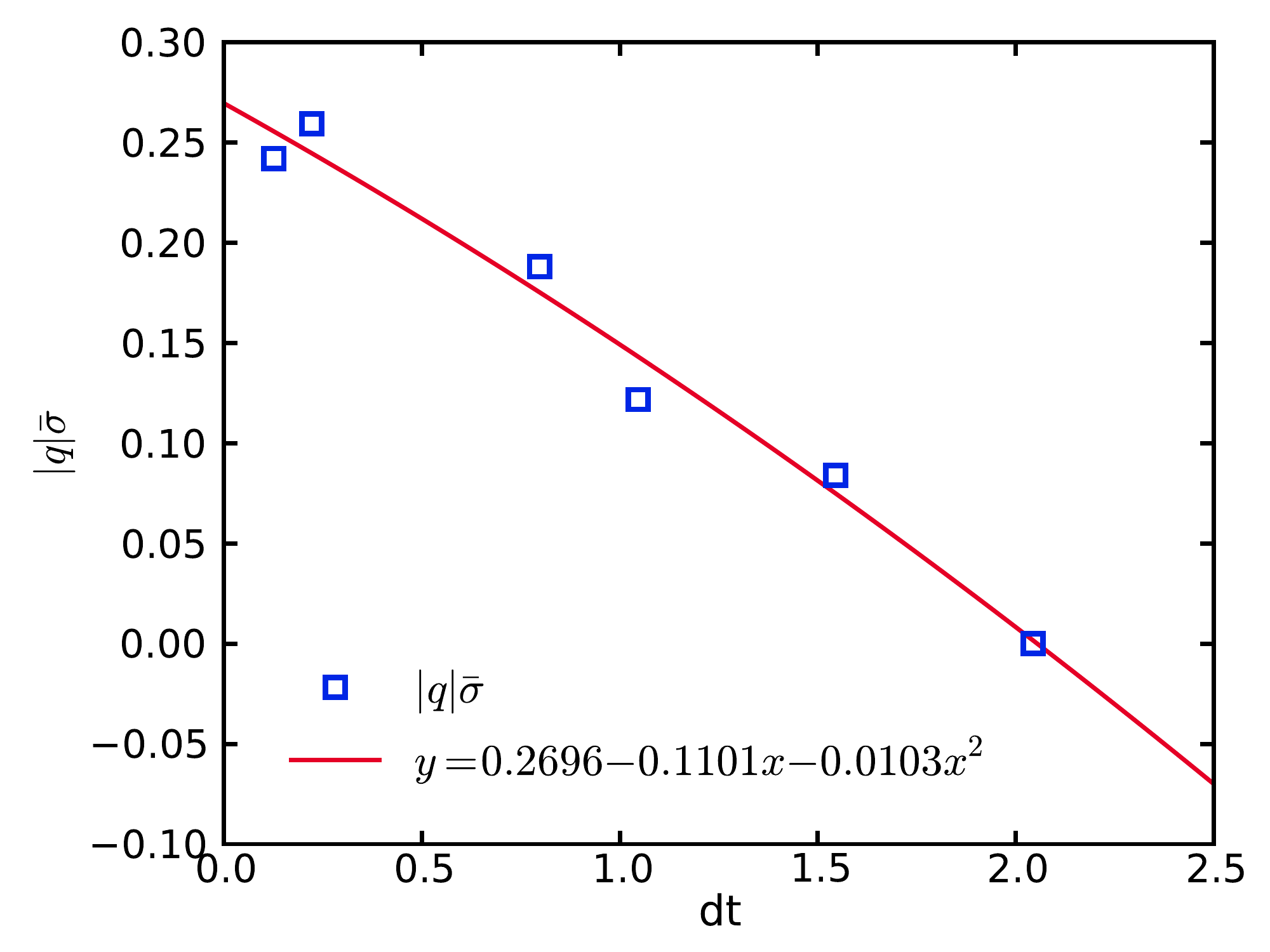}
    \caption{Log-log plot of $\bar{\sigma}=\bar{\sigma}_{q}$ (left), Lin-lin plot of  $|q|\bar{\sigma}_{q}$ (right). EUROSTOXX 50, price date: 01/06/2015}
\end{figure*}

\end{document}